\documentclass[aps,prb,twocolumn,groupedaddress,showpacs]{revtex4-1}
\usepackage{graphicx}
\usepackage{subfigure}
\usepackage{amsmath}
\usepackage{braket}
\usepackage[colorlinks,linkcolor={blue},citecolor={blue},urlcolor={blue}]{hyperref} 
\newcommand{\bA}{\mathbf{A}}
\newcommand{\bB}{\mathbf{B}}
\newcommand{\br}{\mathbf{r}}
\newcommand{\bk}{\mathbf{k}}
\newcommand{\bv}{\mathbf{v}}

\begin{document}

\title{Local versus extended deformed graphene geometries for valley filtering}
%\title{Comparison of deformed graphene geometries  for valley filtering purposes}
%\title{Comparison of valley filter efficiencies for local and extended deformed graphene geometries}
%\title{Local versus extended deformed graphene geometries for valley filtering}
%\title{Local versus extended deformed graphene geometries for production of valley filtered currents}

\author{Dawei Zhai and Nancy Sandler}
\email[]{sandler@ohio.edu}
\affiliation{Department of Physics and Astronomy, Ohio University, Athens, Ohio 45701, USA}

\date{\today}

\begin{abstract}
The existence of two-inequivalent valleys in the band structure of graphene has motivated the search of mechanisms that allow their separation and control for potential device applications. Among the several schemes proposed in the literature, strain-induced out-of-plane deformations  (occurring naturally or intentionally designed in graphene samples), ranks among the best candidates to produce separation of valley currents. Because valley filtering properties in these structures is, however, highly dependent on the type of deformation and setups considered, it is important to identify the relevant factors determining optimal operation and detection of valley currents. In this paper we present a comprehensive comparison of two typical deformations commonly found in graphene samples: local centro-symmetric bubbles and extended folds/wrinkles. Using the Dirac model for graphene and the second-order Born approximation we characterize the scattering properties of the bubble deformation, while numerical transmission matrix methods are used for the fold-like deformations. In both cases, we obtain the dependence of valley polarization on the geometrical parameters of deformations, and discuss their possible experimental realizations. Our study reveals that extended deformations act as better valley filters in broader energy ranges and present more robust features against variations of geometrical parameters and incident current directions.
 
\end{abstract}

\maketitle

\section{Introduction}
\label{SecI}
In many materials, energy bands exhibit  a discrete number of inequivalent local minima or maxima for specific values of momenta, usually known as {\it valleys}, with potential use as quantum numbers to encode, process and carry information\cite{ValleytronicsReview,ViewpointValleytronics,KatsnelsonGrapheneBook}. The field of {\it valleytronics}, i.e. the manipulation of the valley degree of freedom for electronic purposes, has emerged in recent years as an active area of research mainly due to two reasons: 1) The availability of new mono- and few-layer materials that possess two inequivalent valleys at the edges of the Brillouin zone. In some of these structures, these {\it valleys} appear to be relatively easy to access, making them ideal components of a binary variable or pseudo spin. 2) Valley separation may reveal novel physical phenomena that can be exploited in the development of the next generation of electronic devices, e.g. sensors, filters, etc; beyond current semiconductor technologies\cite{valleyNatPhys2007,ValleyPRL2007,ValleytronicsNatNanoZeng,ValleytronicsNatNanoMak,ValleytronicsDiamond}. 

Among the wide variety of materials investigated, graphene and monolayer transition metal dichalcogenides (TMDs) stand out as the most promising candidates for valleytronics\cite{valleyNatPhys2007,ValleyPRL2007,ValleytronicsNatNanoZeng,ValleytronicsNatNanoMak}. These materials have a honeycomb crystal structure that renders two inequivalent energy minima,  labeled $K$ and $K'$, acting as components of a pseudospin degree of freedom in momentum space. For graphene in particular, various schemes have been proposed to achieve {\it valley polarization} (also referred to as {\it valley filtering}), i.e., the generation of a charge current composed of electronic states from only one valley. One of the first proposals, advanced by Rycerz et al., consisted of a sharp constriction within a long ribbon with zigzag edges \cite{valleyNatPhys2007}. In this particular geometry an incident current becomes valley polarized after crossing the constriction. The scheme exploits the very small number of modes present in the constricted region (ideally one or two to obtain maximum efficiency), with a filtering capacity very sensitive to the constriction size as well as to the edge profile of the sample. Interestingly, small constrictions in graphene have revealed very rich physics -such as Coulomb blockade- with properties strongly dependent on substrate materials\cite{Bischoff-KlausQD}, features that  preclude their application as valley polarizers. As extensions of these ideas, several authors proposed a filtering mechanism based on the same group velocities but different band curvatures (effective masses) of states around the two valleys and far away from Dirac points\cite{TrigonalWarpingPRL2008,TrigonalWarpingJPhysCondensMatter}. The effect, known as  'trigonal warping', has the advantage of eliminating the restriction imposed by a small-sized constriction, but  has the drawback of being effective only at large energies. In addition, the degree of valley polarization is very sensitive to the relative orientation of the confined region with respect to crystalline directions, as well as on perfect edge terminations. Other proposals involve defects or crystal dislocations, or mirror symmetry breaking potentials as scattering centers that would result in valley polarization\cite{ValleyLineDefectPRL2011,ValleyLineDefectPRB2014,MahmoudValleyFilter}, as well as the use of polarized light for states in the 'trigonal warping' energy range\cite{TrigonalWarpingLight,LightBilayer}. Experimental realizations of these schemes however, have proven to be quite challenging: line defects have to be atomically controlled over long distances in the first case, while high frequency lasers needed to achieve polarization produce  highly non-equilibrium electron populations, that may relax via plasmon excitations and/or damage samples, thereby introducing unwanted disorder effects\cite{LaserDamage1,LaserDamage2}.

From a practical perspective, it is crucial to maintain the quality of the material in order to exploit its metallic conduction capabilities and thus obtain sizable valley polarized currents. The importance of minimizing disorder effects was specifically demonstrated in non-pristine TMD materials where it was shown that a highly reduced valley polarization was due to inter-valley scattering introduced by impurities\cite{ValleyHallMoS2}. In this regard, graphene has the advantage of featuring a higher crystal quality that ensures longer inter-valley scattering lengths, even at room temperatures\cite{ValleytronicsNatPhysBilayerGraphene}. Clearly, the scientific challenge nowadays resides on finding out simple mechanisms that exploit graphene's quality to produce valley filtered currents in a controllable manner.

Following a traditional approach extensively used in the semiconductor electronic industry\cite{ValleytronicsReview}, mechanical deformations have been advanced as an alternative method to produce valley polarization in graphene. For instance, it has been proposed that a current incident into a region with uniaxial strain can exit with a varying degree of valley polarization dependent on the incident angle with respect to the sample crystalline orientation\cite{ValleyStrainMagnetAPL,ValleyStrainMagnetPRB,ValleyStrainPRL}. 
A few other studies have focused on graphene samples with more complicated but realistic out-of-plane deformations\cite{ValleyStrainACvoltagePRL,TransmissionOutofPlaneDeformationPRL2008}. In all these two-terminal device models, time-reversal symmetry breaking fields, either as periodic time-dependent deformations, externally applied magnetic fields or magnetic materials deposited as barriers beyond the strained region, are necessary to produce the final filtering.
Furthermore, uniform strain profiles on long length-scales and well-defined external magnetic barriers are challenging to achieve in a controllable manner without introducing effects such as strain or spin-orbit coupling, which may provide likely reasons for the absence of experimental implementations of all these proposed devices. 

In this regard, ideal valley filters should take advantage of the exceptional electronic and mechanical properties of graphene, without the need of external magnetic fields or materials, while producing sizable signals whose detection should be relatively easy to achieve. Along these lines, we note that recently, a tight-binding numerical study reported separation of valley currents due to a non-uniform strain produced by a local nanoscale out-of-plane deformation (labeled 'nano-bubble')\cite{Settnes}. In this approach the filtering occurs by spatial separation of valley currents, eliminating the need of some sort of magnetic or time-dependent fields, as required in previous schemes. These authors exploit the existence of a low-energy discrete resonance that enhances the angular separation between the two valley polarized currents. However, the proposal has two important drawbacks: 1) the resonant regime needs to be finely tuned to produce polarization, and 2) as discussed in related works\cite{PeetersBump}, even in these ideal conditions, the transmitted valley current is a rather small fraction of the total incident current and strongly depends on the location and size of the deformation with respect to the contacts, making it potentially hard to detect in available setups. These results bring to light the importance of efficiency not only in the {\it generation} but also in the  {\it detection} of valley filtered currents. 

\begin{figure}[ht]
\includegraphics[width=3.4in]{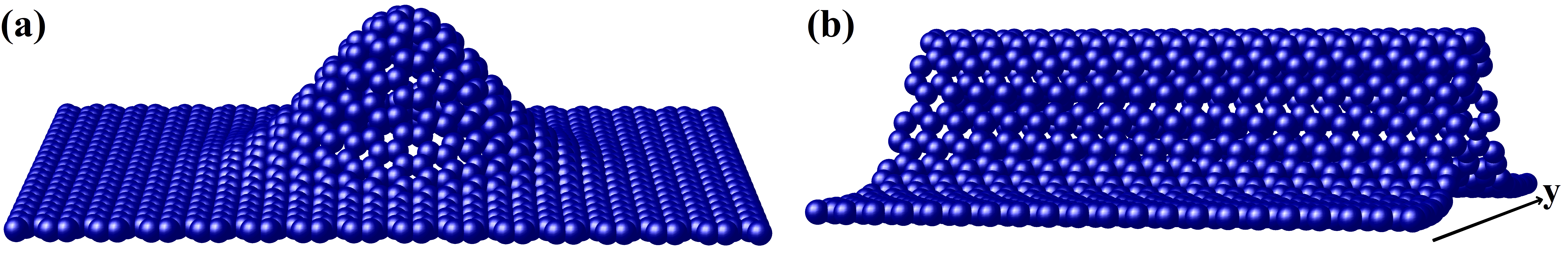}
\caption{Schematics of the out-of-plane Gaussian bump (a) and Gaussian fold (b).}
\label{deformations}
\end{figure}

Following these ideas as guiding principles, we focus on bump-like and fold-like out-of-plane deformations as shown in Fig.~\ref{deformations} that are commonly observed in graphene samples, with the purpose of identifying key parameters that may be used to optimize valley separation for experimental valley current detection. In supported graphene membranes these deformations are usually caused by trapped impurities\cite{Crommie}, deposition on lattice mismatched substrates\cite{Ruoff,FoldPRB2015}, by proper substrate engineering\cite{Eva,Nadya}, or can be produced by appropriate manipulation techniques such as AFM and STM tips\cite{MarkusNanoLett2010,DeformationSTMPRB2012,DeformationSTMNatCommun2014,MarkusNanoLett2017,Annett}. 
We notice in particular that strained graphene with fold-like deformations has already been the subject of experimental transport studies recently\cite{Marc2018}. Thus, the continuous progress in experiments for controllable strained structures, points to the need of determining which geometries and strain profiles optimize valley filtering properties. 

The present study analyzes the transport properties of graphene in the presence of local bubble-like and extended fold-like out-of-plane deformations in terms of the resulting valley polarization of currents incident in the deformed region. The ultimate purpose if to characterize the effectiveness of these structures as valley filters in terms of parameters with experimental relevance in available setups.  As we will show below, our results suggest that, of these two geometries, folds (or equivalent extended non-homogeneous strain geometries) are better valley filter devices in terms of degree of polarization and sizable transmitted currents. The ultimate reason for such improved performance relies on the strong confinement imposed by their structure that also contributes to an optimal spatial separation between valley currents. 

The paper is organized as follows: Section~\ref{SecII} presents the general formalism in terms of the continuum description of strain in graphene, for the two deformations of interest: a centro-symmetric Gaussian bump and an extended Gaussian fold. In Section~\ref{SecIII} we present results for valley polarization obtained with a model for a Gaussian bump based on the continuum Dirac description and standard Lippman-Schwinger scattering methods. In this section, we calculate the corresponding cross sections and introduce a measure of valley polarization based on them. In Section~\ref{SecIV} we discuss the valley polarization produced by the Gaussian fold. In this case, results are obtained from numerical calculation of transmission coefficients, using standard transmission matrix methods. A discussion of possible experimental setups for the detection of valley polarized currents in such systems is proposed in Section~\ref{SecV} together with a summary of results.
 
\section{Continuum field description of strain in graphene}
\label{SecII}

To exploit the continuum description, we take advantage of the low-energy limit of the standard one orbital tight-binding model for electron dynamics in graphene given in terms of two effective Dirac model Hamiltonians for valleys $K$ and $K'$:
\begin{equation}
H_{K, K'}=v_F\boldsymbol{\sigma}\cdot\textbf{p}
\label{H0}
\end{equation}
where the Hamiltonian is written in the valley isotropic basis\cite{ValleyIsotropicNotation}, $v_F\approx10^6$ m/s is the Fermi velocity of pristine graphene, and $\boldsymbol{\sigma}=(\sigma_x,\sigma_y)$ is the Pauli matrix vector.

Strain in graphene samples may have various different origins and in order to incorporate its effects in realistic models it is necessary to establish the specific conditions in which it occurs. For our local bump-like structures, we refer to graphene membranes deposited on top of a locally rough substrate (of natural origin or with a designed pattern to produce artificial 'roughness'\cite{Nadya}) or with an intercalated impurity cluster between graphene and a flat substrate. In the second case of fold-like structures, graphene is positioned on top of an otherwise flat substrate and it is either mechanically or naturally folded\cite{Annett} or wrinkled (due to relaxation of underlying lattice mistmatch induced strains), with the length of these structures much longer than their respective widths. Alternatively, it may be deposited on top of carefully designed substrates where folds form under deposition\cite{Eva}. In the bump scenario, the membrane bends out of plane to accommodate to the roughness of the underlying surface while in the fold, it bends as it folds or wrinkles.  A natural mathematical description of this situation is given by the Monge representation, that refers to a one-to-one mapping between a continuum curved surface and the (flat) Euclidean plane, i.e., $h = h(x,y)$ where $h$ represents the height of the membrane on top of the flat plane\cite{DNelsonbook}. The description is suitable for the situations described above, as the presence of the substrate imposes a firm constraint on the membrane, that impedes longitudinal stretches beyond the deformed region\cite{LandauElasticityBook}. 

Within this approach, it has been shown that the two most important effects produced by deformations involve: 1) changes in the local charge distribution, and 2) modifications in the local hopping parameters\cite{Ando,Manes,CastroNeto2009,curvatureeffect,Fermivelocity}. These are included in the continuum theory as scalar $\Phi(\br)$ and pseudo-vector $\textbf{A(r)}$ potentials\cite{Ando,Manes,GaugeFieldPhysRep}, defined by:
\begin{eqnarray}
\Phi(\br) &= & g_s (\epsilon_{xx} + \epsilon_{yy}) \\
\bA(\br) &=&-\frac{\hbar \beta}{2ae}\left(\epsilon_{xx}-\epsilon_{yy},-2\epsilon_{xy}\right),
\label{eq:fields}
\end{eqnarray}
where $\epsilon_{ij}$ with $i,j=\{x,y\}$ are the components of the strain tensor, $g_s \approx 3$\,eV \cite{gsGuinea2010,Salvador2013,Martin,AlexCroy} and $\beta\approx3$\cite{Ando,GaugeFieldPhysRep,VictorPRB2009,gsGuinea2010,Salvador2013,Martin,AlexCroy} are the corresponding coupling constants. $a$ and $e$ are the lattice constant and magnitude of the electron charge respectively. The strain tensor components
are given by $\epsilon_{ij}=\frac{1}{2}(\partial_j u_i+\partial_i u_j+\partial_i h \partial_j h)$, where $u_{i,j}$ and $h$ are in-plane and out-of-plane displacement fields, respectively\cite{LandauElasticityBook}. Time reversal symmetry imposes opposite signs for the pseudo-vector field at $K$ and $K'$ valleys. As a consequence, the low energy continuum model that includes the effect of deformations is written as:
\begin{equation}
H_{\tau}=v_F\boldsymbol{\sigma}\cdot\left(\textbf{p}+\tau e \textbf{A}\right) + \Phi(\br) \, \sigma_0,
\label{eq:hamilt}
\end{equation}
where $\tau=\pm$ labels each valley, and $\sigma_0 = {\cal{I}}_{2\times2}$ is the identity matrix. For the specific experimental realizations described above, the two types of deformations are modeled by:
\begin{equation}
h(\br)=
\begin{cases}
h_0e^{-r^2/b^2}&\text{Gaussian bump}\\
h_0e^{-y^2/b^2}&\text{Gaussian fold}
\end{cases}.
\end{equation}
In these deformed structures, the linear terms due to the in-plane displacements in the strain tensor are neglected, while those result from the out-of-plane displacement are retained, consistent with expected experimental constraints.

The centrosymmetric local structure of the Gaussian bump shown in Fig.~\ref{deformations}(a) produces scalar and vector potentials given by
\begin{eqnarray}
\bA(\br)&=&-\frac{\overline{\beta}\eta^2}{e v_F}g\left(\frac{r}{b}\right)\left(\cos2\theta,-\sin2\theta\right) \\
\Phi(\br) &=& g_s \eta^2 g\left(\frac{r}{b} \right). 
\label{eq:bumpfields}
\end{eqnarray}

Analogously, the extended Gaussian fold is translationally invariant in $\hat{x}$, chosen along the zigzag crystalline direction as shown in Fig.~\ref{deformations}(b), and producing strain-induced potentials given by
\begin{eqnarray}
\bA(\br)&=&\frac{\overline{\beta}\eta^2}{e v_F}g\left(\frac{y}{b}\right)\left(1,0\right)\\
\Phi(\br) &=& g_s \eta^2 g\left(\frac{y}{b}\right).
\label{eq:foldfields}
\end{eqnarray}

\begin{figure}[ht]
\includegraphics[width=3.4in]{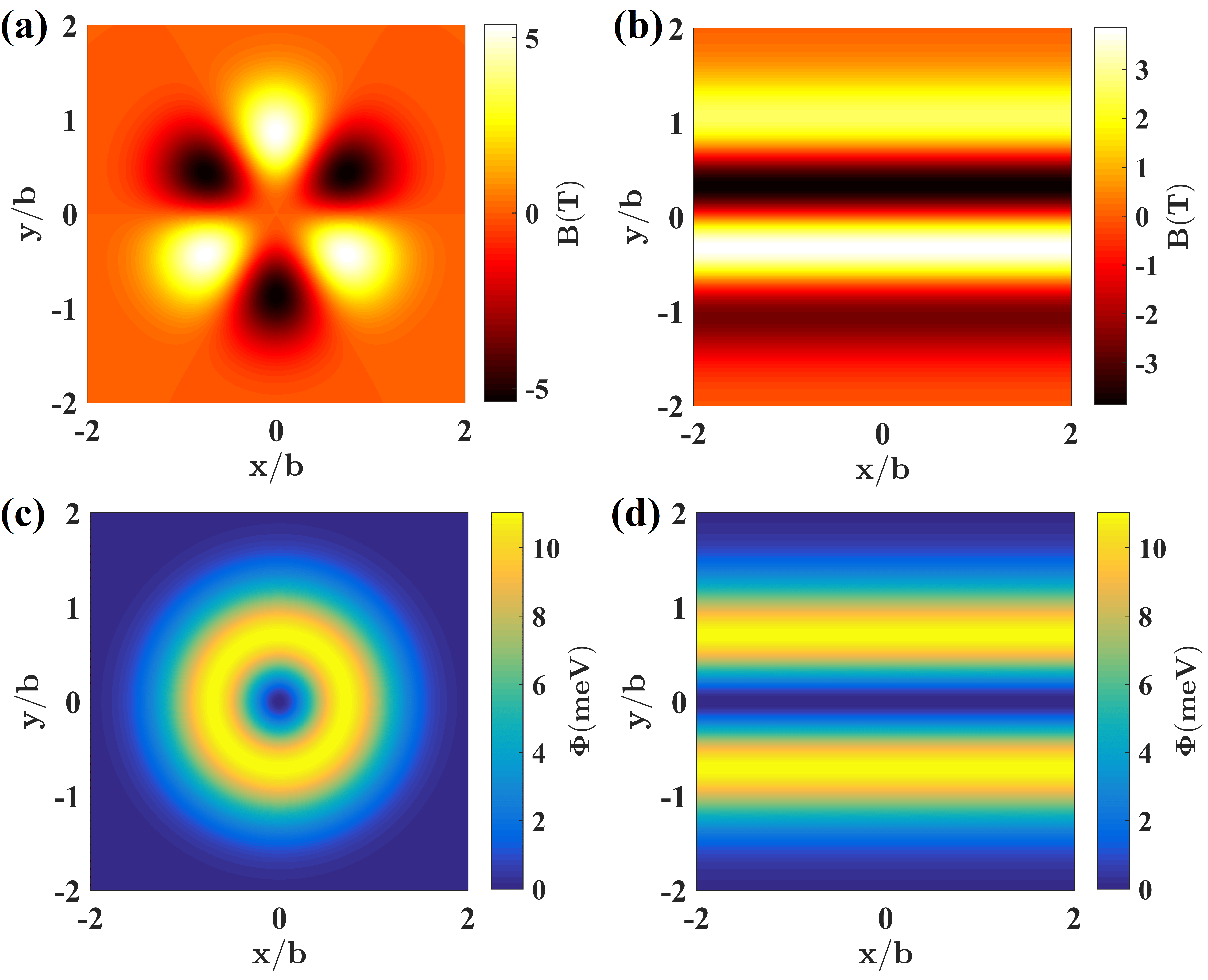}
\caption{Strain induced pseudo-magnetic field $\textbf{B}$ and $\Phi(\br)$ for bump and fold. $\bB$ for bump (a), and fold (b). $\Phi(\br)$ for bump (c) and fold (d). Parameters: $\eta=0.1$, and $b=15$nm.}
\label{Fig2}
\end{figure}

In these expressions $\overline{\beta}=\frac{\hbar\beta v_F}{2ae}\approx7eV$, $g(z)=2z^2e^{-2z^2}$, $\theta$ is the polar angle measured with respect to the zigzag crystalline orientation, and $\eta=\frac{h_0}{b}$ is a measure of the strain strength. 
Fig.\ref{Fig2} shows typical profiles of scalar potential and pseudo-magnetic field $\textbf{B}=\nabla\times\textbf{A}$ for $K$ valley produced by these two deformations. Results for the pseudo-magnetic field $\textbf{B}$ at the $K'$ valley are obtained by the exchange of positive and negative regions. In the following sections, we present results for Gaussian folds along the zigzag direction, while those for a generic orientation of the fold axis are given in Appendix~\ref{App_C}.

\section{Local Gaussian bump deformation}
\label{SecIII}
Effects of a gaussian bump deformation on the electronic properties of graphene have been extensively studied by several authors. Here we briefly review their main findings: Aharonov-Bohm interferences are predicted to occur due to the presence of strain induced pseudo-magnetic fields\cite{ABinterference}. A redistribution of both sublattices' electronic charge densities in equilibrium was predicted\cite{PeetersSublatticeSymmetryBreaking,Salvador2013,Ramon,Martin}, which was afterward confirmed by experimental observation and quantitative characterization of triangular patterns in local STM images of tip-lifted graphene on $\text{SiO}_2$\cite{MarkusNanoLett2017}. The sublattice symmetry breaking phenomena was also confirmed on samples with deformations produced by impurities intercalated between the graphene membrane and its substrate (hBN)\cite{MarkusNanoLett2017}. Furthermore, the bump deformation has also been predicted to bend or focus currents\cite{SzpakNJP}, and even produce valley polarization under 'resonant conditions' for deformations with strong scattering potentials\cite{Settnes,PeetersBump}.

In order to asses the efficiency of this geometry towards the production of valley polarized currents and evaluate conditions for its detection, we describe in detail its scattering effects on an electron current composed of states from both valleys. 

\subsection{Scattering within Born approximation formalism}
\label{ref:SubsecScatBorn}

We evaluate the scattered wave function $\Psi_{\bk}(\br)$, using the Lippmann-Schwinger equation as
\begin{equation}
\Psi_{\bk}(\br)=\psi_{\bk'}(\br)+\int G(\br,\br')V(\br')\Psi_{\bk}(\br')d\br',
\end{equation}
where $\psi_{\bk'}(\br)$ is the incident wave function, $G(\br,\br')$ the Green's function, $V(\br')$ the scattering potential due to the deformation, and the integration is over the deformed region. Within the first order Born approximation, the scattered wave function is approximated as 
\begin{equation}
\Psi^{(1)}_{\bk}(\br)=\psi_{\bk'}(\br)+\int G(\br,\br')V(\br')\psi_{\bk'}(\br')d\br',
\label{1stBorn}
\end{equation}
from which the second order is obtained by the standard procedure:
\begin{equation}
\Psi^{(2)}_{\bk}(\br)=\psi_{\bk'}(\br)+\int G(\br,\br')V(\br')\Psi_{\bk}^{(1)}(\br')d\br'.
\label{2ndBorn}
\end{equation}
Herein, $\br$ refers to the point of evaluation of $\Psi_{\bk}(\br)$, $\bk'$ and $\bk$ label incident and scattered wave vectors, respectively. In the following, we will consider particle states with the same group velocity $\left(\bv=\frac{1}{\hbar}\frac{\partial E}{\partial \bk}\right)$ irrespective from its valley origin. With these considerations, the incident wave function reads
\begin{equation}
\psi_{\bk'}(\br)=u_{\bk'}e^{i\bk' \cdot \br}
=\frac{1}{\sqrt{2}}
\begin{pmatrix}
e^{-i\theta'/2}\\e^{i\theta'/2}
\end{pmatrix}
e^{i\bk' \cdot \br},
\end{equation}
where $\bk'=k'e^{i\theta'}$. The Green's function is given by
\begin{equation}
G(\br,\br')=-\frac{ik}{4\hbar v_F}
\begin{pmatrix}
H_0^{(1)}(k\rho)&ie^{-i\tilde{\theta}}H_1^{(1)}(k\rho)\\
ie^{i\tilde{\theta}}H_1^{(1)}(k\rho)&H_0^{(1)}(k\rho)
\end{pmatrix},
\end{equation}
where $H_n^{(1)}(k\rho)$ is the first kind Hankel functions of order $n$, and $\boldsymbol{\rho}=\br-\br'=\rho e^{i\tilde{\theta}}$.\cite{Greenfunction}
Following Eq.~\ref{eq:hamilt}, the scattering potential contains the contribution of the scalar field $\Phi(\br)$ proportional to the identity matrix and that of the pseudo vector potentials that reads
\begin{equation}
V^{\tau}(\br')=-\tau\overline{\beta}\eta^2g\left(\frac{r'}{b}\right)
\begin{pmatrix}
0&e^{i2\phi}\\
e^{-i2\phi}&0
\end{pmatrix},
\end{equation}
where $\phi$ is the polar angle of $\br'$.

In a scattering experiment, the detector is usually placed far away from the scatterer, i.e. $r\rightarrow\infty$. In this case, one can make the approximations $\tilde{\theta}\approx\theta$ and $\rho\approx r-\hat{r}\cdot \br'$, where $\hat{r}$ is the unit vector along $\br$. Using the asymptotic expression of the Hankel function $H_n^{(1)}(z)\rightarrow\sqrt{\frac{2}{\pi z}}e^{iz}e^{-i(\frac{n}{2}\pi+\frac{\pi}{4})}$ as $z\rightarrow\infty$, one can then easily verify that
\begin{equation}
\begin{aligned}
G(\br,\br')\approx&
-\sqrt{\frac{ik}{2\pi r\hbar^2v^2_F}}\braket{\br|\psi_{\bk}}\braket{\psi_{\bk}|\br'}\\
=&-\sqrt{\frac{ik}{2\pi\hbar^2v^2_F}}\frac{e^{ikr}}{\sqrt{r}}u_{\bk}\left(u_{\bk}\right)^\dagger e^{-i\bk\cdot\br'}
\end{aligned},
\label{Green@infty}
\end{equation}
where we have used the fact that $\bk=ke^{i\theta}=k\hat{r}$\cite{BumpJAP2012}.

We first focus on the scattering produced by the pseudo vector potential $V^{\tau}(\br')$, the effects of the scalar field $\Phi(\br)$ will be discussed later. From Eqs.~\ref{1stBorn}, \ref{2ndBorn}, and \ref{Green@infty} one obtains
\begin{equation}\label{2nd_scattered_wave}
\Psi^{(n),\tau}_{\bk}(\br)=\psi_{\bk'}(\br)+f^{(n),\tau}(\theta,\theta')\frac{e^{ikr}}{\sqrt{r}}u_{\bk},
\end{equation}
where $n = 1, 2$ indicates the order of the expansion, and the corresponding form factors are given by:
\begin{equation}
\begin{aligned}
f^{(1),\tau}(\theta,\theta')&=-\sqrt{\frac{ik}{2\pi\hbar^2v^2_F}}V^{\tau}_{\bk,\bk'}\\
f^{(2),\tau}(\theta,\theta')&=f^{(1),\tau}(\theta,\theta')-\sqrt{\frac{ik}{2\pi\hbar^2v^2_F}}(VGV)^{\tau}_{\bk,\bk'}
\end{aligned}
\end{equation}
with the shorthand notations
\begin{equation}
\begin{aligned}
V^{\tau}_{\bk,\bk'}&=\braket{\psi_{\bk}(\br')|V^{\tau}(\br')|\psi_{\bk'}(\br')}\\
(VGV)^{\tau}_{\bk,\bk'}&=\braket{\psi_{\bk}(\br')|V^{\tau}(\br')G(\br',\br'')V^{\tau}(\br'')|\psi_{\bk'}(\br'')}
\end{aligned}
\end{equation}

The differential cross section for each valley can be obtained from the form factor as
\begin{equation}
\sigma^{(n),\tau}_D(\theta,\theta')=\left|f^{(n),\tau}(\theta,\theta')\right|^2.
\end{equation}
One can verify that
\begin{equation}
\begin{aligned}
V^{+}_{\bk,\bk'}&=-V^{-}_{\bk,\bk'}\\
(VGV)^{+}_{\bk,\bk'}&=(VGV)^{-}_{\bk,\bk'}
\label{eq:identities}
\end{aligned},
\end{equation}
The explicit expressions for these quantities are given in Appendix \ref{App_A}. Due to these identities, results for valley $K'$ can be written in terms of the quantities for valley  $K$. In the following discussion, expressions for $V_{\bk,\bk'}$ and $(VGV)_{\bk,\bk'}$ refer to $K$ valley.

Identities in Eq.~\ref{eq:identities} clearly reveal identical cross sections for both valleys at first order but different ones at 2nd order (in the absence of scalar scattering) since:
\begin{equation}
\begin{aligned}
\sigma_D^{(2),+}&=\frac{k}{2\pi\hbar^2v_F^2}\left|V_{\bk,\bk'}+(VGV)_{\bk,\bk'}\right|^2\approx\sigma_D^{(1)}+\frac{1}{2}\Delta\\
\sigma_D^{(2),-}&=\frac{k}{2\pi\hbar^2v_F^2}\left|-V_{\bk,\bk'}+(VGV)_{\bk,\bk'}\right|^2\approx\sigma_D^{(1)}-\frac{1}{2}\Delta
\end{aligned}
\label{eq:sigmasforKK'},
\end{equation}
where
\begin{equation}
\begin{aligned}
\Delta=\sigma_D^{(2),+}-\sigma_D^{(2),-}=\frac{2k}{\pi\hbar^2v_F^2}\Re\left[V_{\bk,\bk'}\cdot (VGV)_{\bk,\bk'}\right]
\end{aligned}
\end{equation}
is the difference between the differential cross sections of the two valleys. $\Re$ represents the real part. Note that $\Delta$ scales with the strength intensity as $\eta^6$. 

In order to quantify the degree of valley filtering, we use a standard definition of the angle dependent polarization coefficient $P$, given by:
\begin{equation}
P=\frac{\sigma_D^{(2),+}-\sigma_D^{(2),-}}{\sigma_D^{(2),+}+\sigma_D^{(2),-}}\approx\frac{\Delta}{2\sigma_D^{(1)}}.
\end{equation}
This expression indicates that increased polarization can be achieved by larger strain intensities as $P \propto \eta^4$.

\subsection{Results and discussion}
\begin{figure}[ht]
\centering
\includegraphics[width=3.4in]{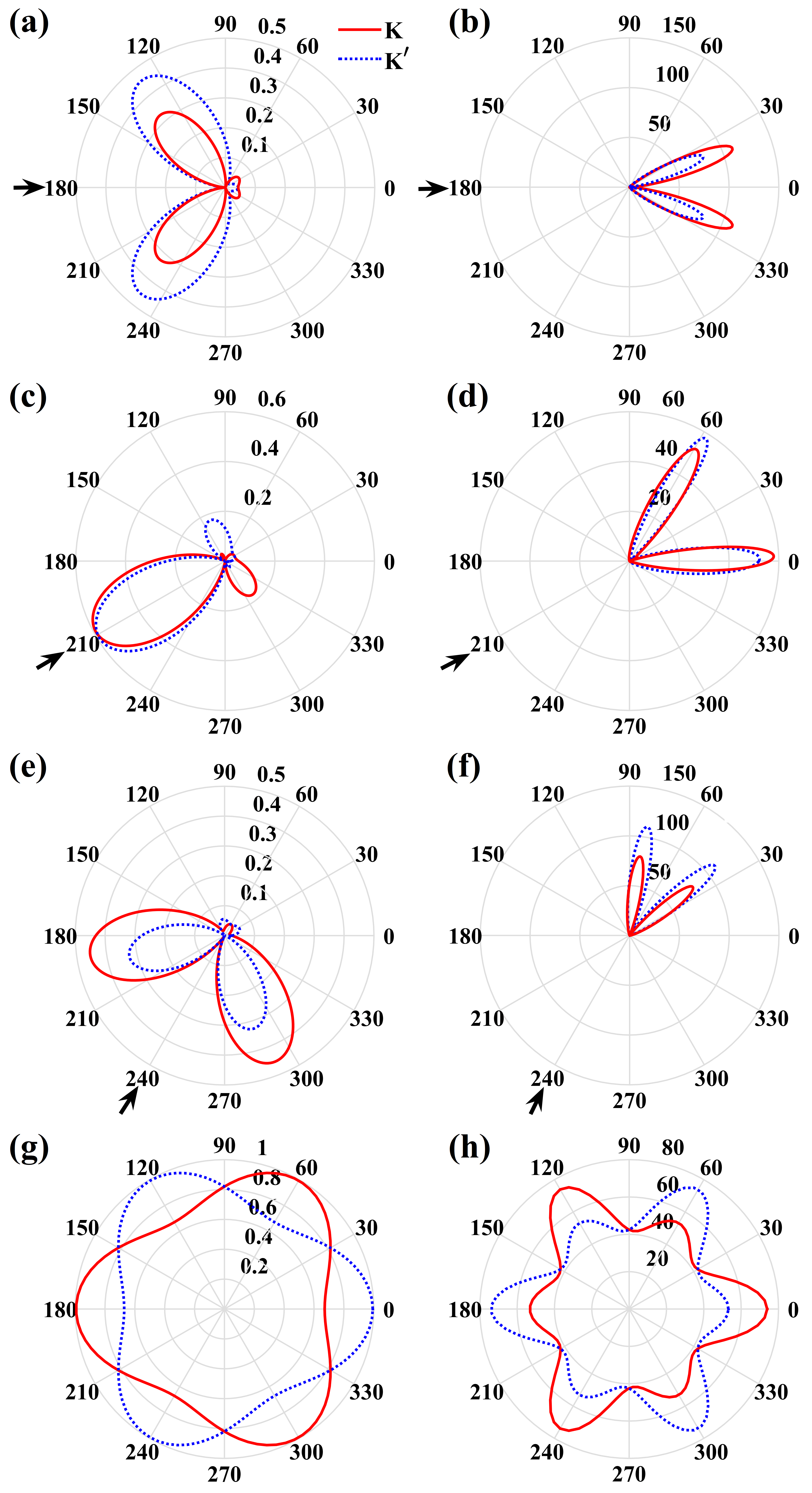}
\caption{Polar plots of the differential cross section (\AA) for $K$ (red solid) and $K'$ (blue dotted) valleys with different incident angles indicated by the arrows and total cross section as function of incident angles. (a-b) $\theta'=0^\circ$, (c-d) $\theta'=30^\circ$, (e-f) $\theta'=60^\circ$. (g-h) shows the total cross section versus the incident angle $\theta'$. Left column corresponds to energy $E=20$meV, while right column corresponds to energy $E=300$meV. Other parameters: $b=15$nm, $\eta=0.1$, $E_b \approx 44$meV.}
\label{cross_section_vector_only}
\end{figure}

We begin by presenting numerical results for the differential cross section due to the pseudo-vector potential. The conditions for convergence for the series expansion are different in the low- and high-energy regimes defined by $kb\ll1$ (or $E=\hbar v_F k \ll E_b$), and $kb\gg1$ (or $E=\hbar v_F k \gg E_b$) respectively.  Here $E_b=\hbar v_F/b$ is the natural energy scale of the potential associated with the width of the deformation $b$, playing the role of the scattering length.
In the low energy regime, convergence is assured as long as  $V_{max}\ll E \ll E_b$, with  $V_{max} = \bar{\beta} \eta^2/2.71828..$. Equivalently, in the high energy regime, convergence will occur for all energies satisfying $E \gg E_b \gg V_{max}$\cite{sakurai2011modernQM}. Fig.~\ref{cross_section_vector_only} shows results in the low and high energy regimes in the left and right columns respectively, for an electronic state with incident momentum $\bk$.

The figures show that the degree of filtering is highly dependent on the incident direction, with maximum polarization (considering all outgoing directions) occurs for the zigzag crystalline directions (integer multiples of $60^\circ$) in both low and high energy regimes, as clearly shown in Fig.~\ref{cross_section_vector_only} (g, h) where the results of total cross section versus the incident angles are presented. Notice that the polarization is reversed as the incident angle changes by $60^\circ$, e.g. first and third rows of Fig.~\ref{cross_section_vector_only}, reflecting the changes in the underlying pseudo-magnetic field pattern as the rotation is carried out \cite{RotationEffectStrainPRB2016}. This also suggests that the magnitude of the polarization can be switched by properly controlling the incident direction. Overall, the filtering is more effective at high energies, giving bigger differential cross sections. Notice however that currents for both valleys coexist in the same spacial region making the scattered currents only partially polarized.

\begin{figure}[ht]
\includegraphics[width=3.4in]{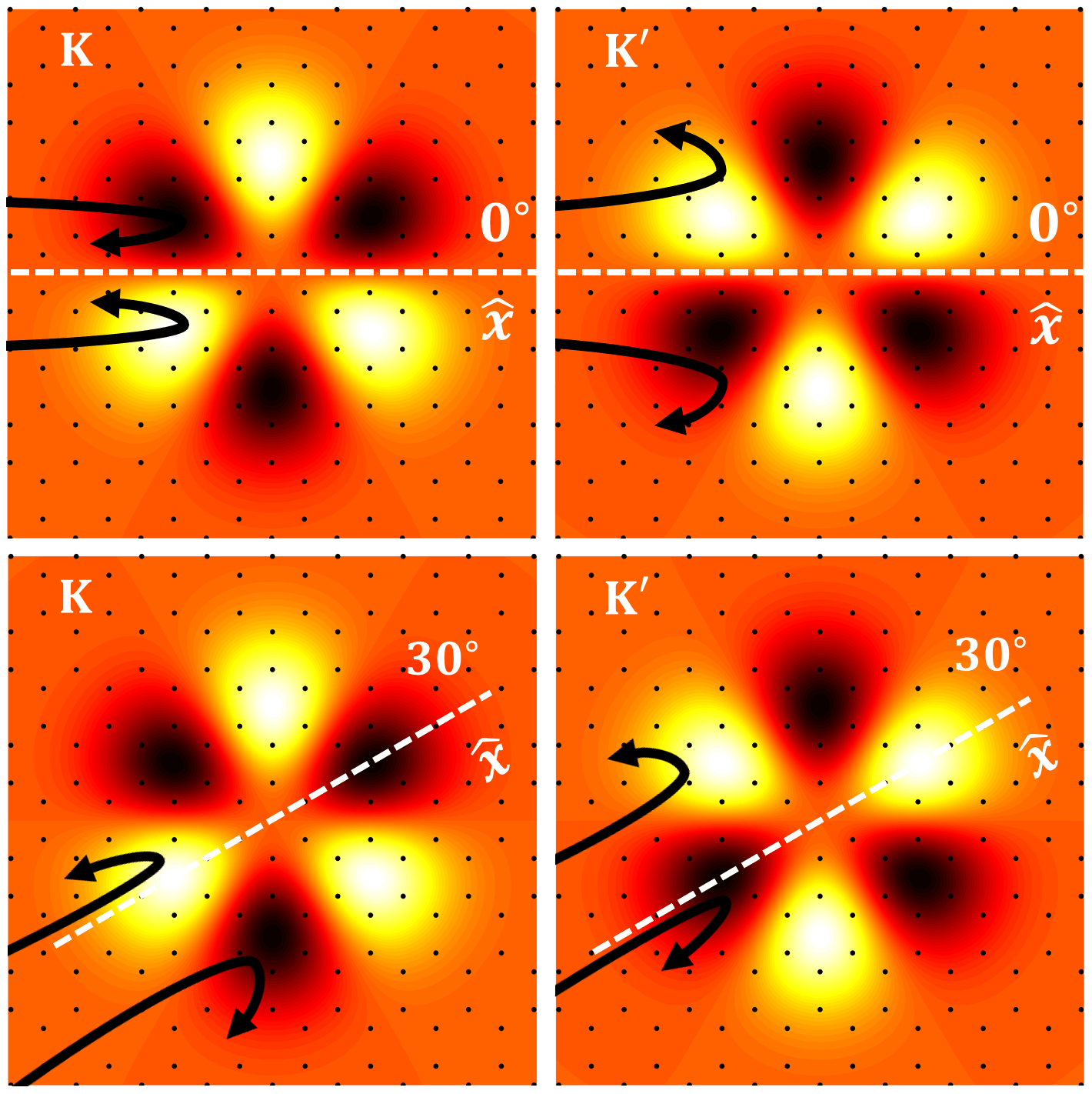}
\caption{Schematics of classical trajectories of electron motion in different magnetic field distributions. Top (bottom) panel corresponds to antisymmetric (symmetric) field distribution with respect to the incident direction, which is chosen as $\hat{x}$. Left (right) column corresponds to $K$ ($K'$) valley.}
\label{classical_trajectory}
\end{figure}

The profile of the differential cross section can be intuitively understood by considering the classical motion of electrons in a magnetic field as depicted in Fig.~\ref{classical_trajectory}. The pseudomagnetic field $B\propto\frac{4\overline{\beta}\eta^2}{ev_Fb}$ renders a magnetic length $l_B=\sqrt{\frac{\hbar}{eB}}=b^{1/2}\sqrt{\frac{\hbar v_F}{4\overline{\beta}\eta^2}}$. Electrons in the low (high) energy regime with $kb\ll 1\,(\gg 1)$ have incident energy much smaller (larger) than the energy associated with the mangetic field ($\propto\hbar v_F/l_B$), thus are more likely to be reflected (transmitted). Note also that the symmetric features of the differential cross section are consistent with the underlying distribution of the pseudo-magnetic field:  the scattering cross section for each valley is symmetric with respect to an axis going through the center of the bump along the incident direction, i.e. $\theta'=2n\times30^\circ$ ($n$ integer), when the distribution of the pseudo-magnetic field is antisymmetric (1st and 3rd rows of Fig.~\ref{cross_section_vector_only}). This results from a change in sign of the pseudo-magnetic fields at the two sides of the axis that bend the electron's trajectory by equal amounts but opposite directions. The first panel of Fig.~\ref{classical_trajectory} shows schematics of classical electron trajectories in the presence of an anti-symmetric distribution of pseudo-magnetic field with respect to the symmetry axis $\hat{x}$ set by the incident direction. For a given valley, the magnetic field satisfies $B(x,-y)=-B(x,y)$ and electron motion is completely opposite at $(x,y)$ and $(x,-y)$. This results into a symmetric scattering cross section profile with respect to $\hat{x}$. For a more general case where the pseudo-magnetic field distribution is not anti-symmetric with respect to the incoming direction, the scattering cross section is asymmetric (middle row of Fig.~\ref{cross_section_vector_only}).
 
The origin of valley polarization effect can also be understood by comparing the two panels in the first row of Fig.~\ref{classical_trajectory}. The pseudo-magnetic field, which is non-uniform in space, exhibits opposite signs in the two valleys, i.e. $B_K(x,y)=-B_{K'}(x,y)$. Consequently, incident electrons with the same initial conditions from the two valleys undergo different motions, not only with opposite directions, but also along different paths. This is the ultimate origin of the valley polarization phenomena. 

Finally, for incident angles $\theta'=(2n+1)\times30^\circ$, the pseudo-magnetic field is symmetric with respect to an axis crossing through the center of the bump along $\theta'$ (shown for $n=0$ in Fig.~\ref{classical_trajectory}). As a consequence, the scattering cross sections of the two valleys are symmetric with respect to each other (middle row in Fig.~\ref{cross_section_vector_only}). In this case, the pseudo-magnetic field satisfies $B(x,y)=B(x,-y)$ for a given valley, and $B_K(x,y)=-B_{K'}(x,-y)$ between the valleys. Thus the path followed by an incident electron from valley $K$ through $(x,y)$ is the same as the path followed by an electron from valley $K'$ through $(x,-y)$ but on opposite directions. This renders the symmetric scattering cross section profiles shown in Fig.~\ref{cross_section_vector_only}(c,d), thus the two valleys exhibit identical total cross section as shown in Fig.~\ref{cross_section_vector_only} (g, h).

\begin{figure}[ht]
\centering
\includegraphics[width=3.4in]{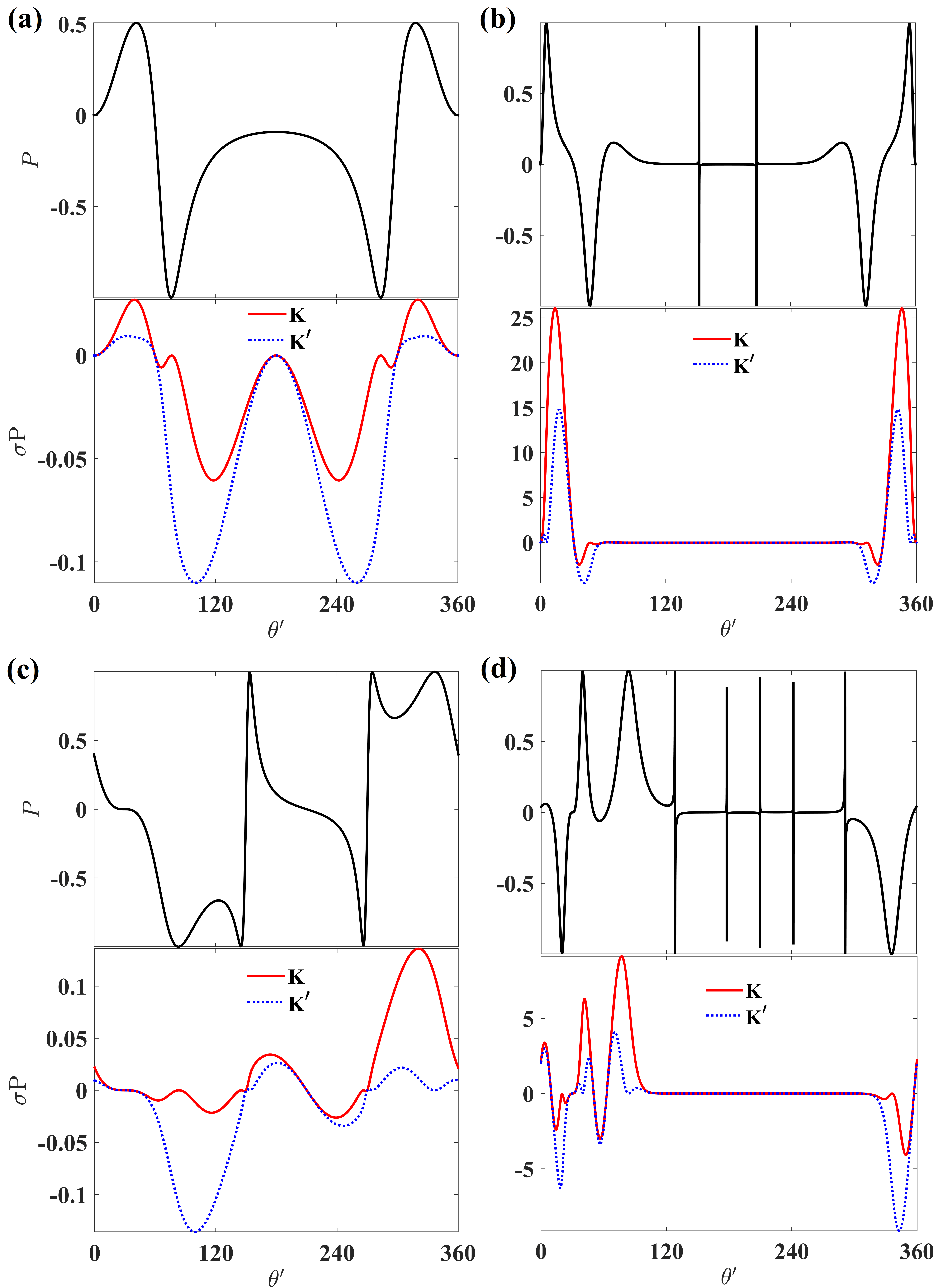}
\caption{Polarization $P$ (black), $\sigma_D P$ for $K$ (red solid) and $K'$ (blue dotted) valley corresponding to panels (a-d) in Fig.~\ref{cross_section_vector_only}. The spikes in polarization plots correspond to directions with zero values of the second order correction, around which abrupt change of sign in the polarization occur as the angle of incidence is changed}
\label{P_and_SigmaP}
\end{figure}

To quantify the valley polarization effect, we calculate the values for $P$ and the product $\sigma_D P$ as a measure of the intensity of polarized currents. Fig.~\ref{P_and_SigmaP} shows the results corresponding to panels (a-d) of Fig.~\ref{cross_section_vector_only}. In the low energy regime (left column, black curves) large polarization values can be obtained in wide angular regions. However, the corresponding differential cross section has small magnitude, indication of weak scattering. Therefore, the product $\sigma_D P$, used to estimate the amount of detected current, is small (red solid and blue dotted curves). In the high energy regime (right column, Fig.~\ref{P_and_SigmaP}), the differential cross section reaches substantial values, however, the scattering events are confined to very narrow angular regions around the incident direction. Consequently, both $P$ and $\sigma_D P$ show narrow peaks and dips in the high energy regime. Note that the highly singular polarization peaks shown in panels (b) and (d) are a consequence of the abrupt change of sign in the polarization as the angle of incidence is changed. The narrow angular distribution and the close proximity to the incident direction impose serious difficulties for the detection of the scattered current:  a detector (or contacts) with high angular resolution is required, and the incident current, usually fully unpolarized, is likely to overwhelm the weakly polarized currents. 

\begin{figure}[ht]
	\centering
	\includegraphics[width=3.4in]{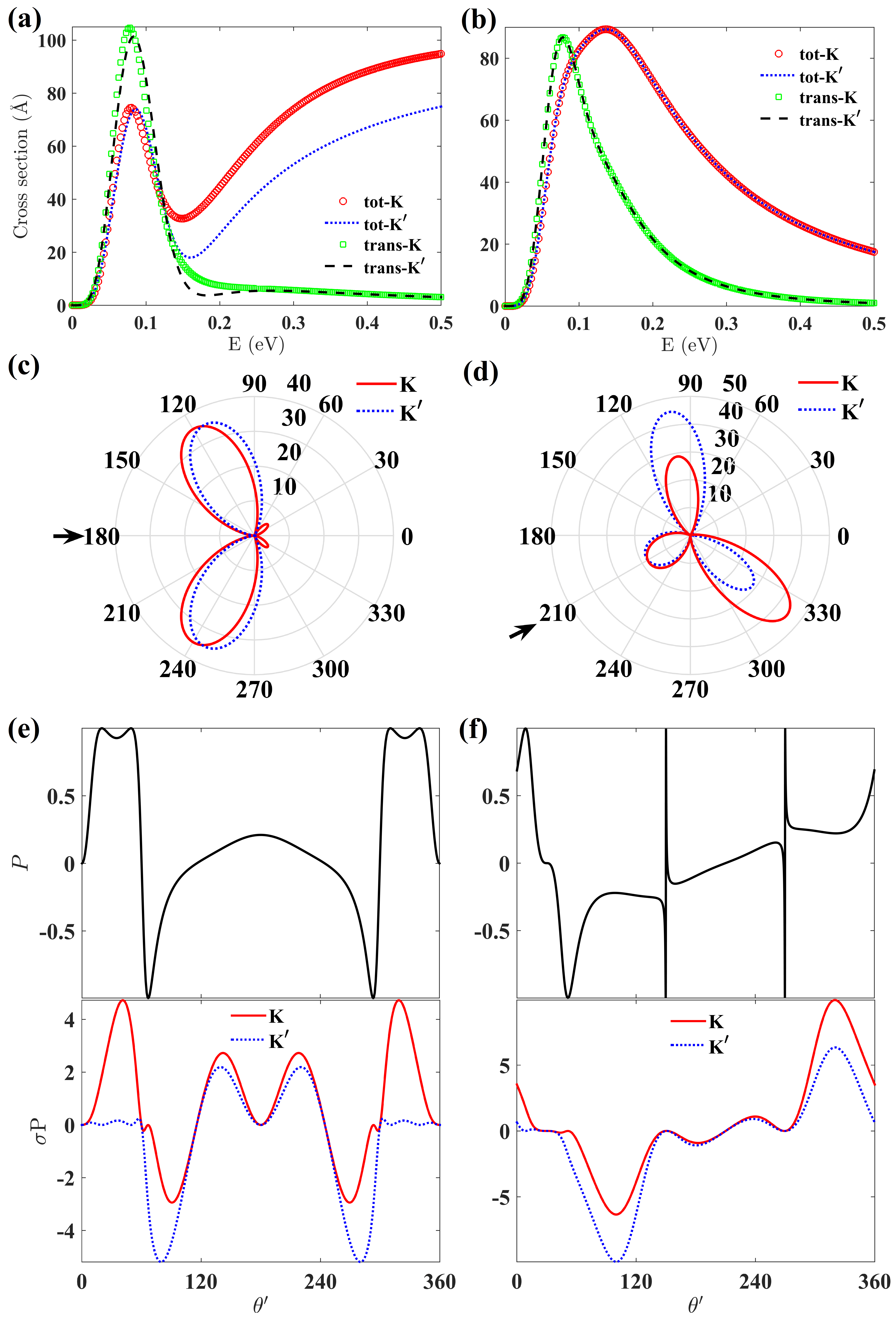}
	\caption{$\sigma_{tot}$ and $\sigma_{trans}$ vs incident energy $E$ for incident angle (a) $0^{\circ}$ and (b) $30^{\circ}$. (c) Differential cross section for $0^{\circ}$  incidence with $E=80$ meV.  (d) Differential cross section for $30^{\circ}$ incidence with $E=93$ meV. (e) Polarization $P$ and $\sigma_D P$ corresponding to data in (c). (f) Polarization $P$ and $\sigma_D P$ corresponding to data in (d). }
	\label{Tot_and_trans}
\end{figure}

To identify the optimal conditions for generation and detection of valley polarization in these geometries, total $\sigma_{tot}(\theta')=\int_{0}^{2\pi}\sigma_D(\theta, \theta') d\theta$ and transport cross sections $\sigma_{trans}(\theta')=\int_{0}^{2\pi}[1-\cos(\theta'-\theta)]\sigma_D(\theta, \theta') d\theta$ are calculated for different incident directions $\theta'$ (measured with respect to the crystalline orientation). Transport cross sections are maximum if both $\left[1 - \cos(\theta'-\theta)\right]$ and $\sigma_D$ are simultaneously large, i.e. if $\sigma_D$ is large in the direction perpendicular to the incident direction ($\theta' - \theta \simeq 90^{\circ}$) and vanishing otherwise. Panels (a) and (b) in Fig.~\ref{Tot_and_trans} shows results for $\sigma_{tot}(\theta)$ and $\sigma_{trans}(\theta)$ vs incident energy for two different incident angles $\theta$. Note that $\sigma_{tot}$ exhibits resonance peaks at energies (e.g. around $80-150$ meV) within the validity of the linear dispersion represented by the Dirac model, and considerably smaller than the high energy used in Fig.~\ref{cross_section_vector_only}. Similarly, $\sigma_{trans}$ shows resonances within this same energy regime, confirming that the cross section is largest at near perpendicular directions with respect to the incident direction. Panels (c) and (d) in Fig.~\ref{Tot_and_trans} show differential cross sections for energies in the resonance regime for both incident directions. Strong scattering is observed for $\theta = 30^{\circ}$ with values for the differential cross section comparable to those in the high energy regime in Fig.~\ref{cross_section_vector_only}. This would facilitate the detection of polarized scattering current in experimental settings since the contribution of the unpolarized incident current can be avoided. Finally, panels (e) and (f) of Fig.~\ref{Tot_and_trans} show the results of $P$ and $\sigma_D P$ corresponding to panels (c) and (d) above.  $\sigma_D P$ exhibits large amplitudes over a wide angular range, indicating that strong scattering and high polarization can be achieved simultaneously within a reasonably wide region. Comparison of results for incident angles $\theta=0^{\circ}$ and $\theta=30^{\circ}$, suggests that the former case is more promising due to wider regions with vanishing values for $\sigma_D P$ for one of the two valleys, but with the disadvantage of a narrow angular amplitude close to the incident direction.

Up to this point the scattering effects of the scalar potential $\Phi$ has been discarded. By itself, the scalar potential cannot give rise to valley filtered currents due to its valley-independent nature. However when combined with the pseudo-vector potential $V^\tau$, it renders a total scattering potential equal to $\Phi+V^{\tau}$. Because the differential cross section is given by the form factor squared, it is clear from this expression that the two valleys will display different behavior, already at first order due to the opposite signs of $V^\tau$ for the two valleys. 

\begin{figure}[ht]
	\includegraphics[width=3.4in]{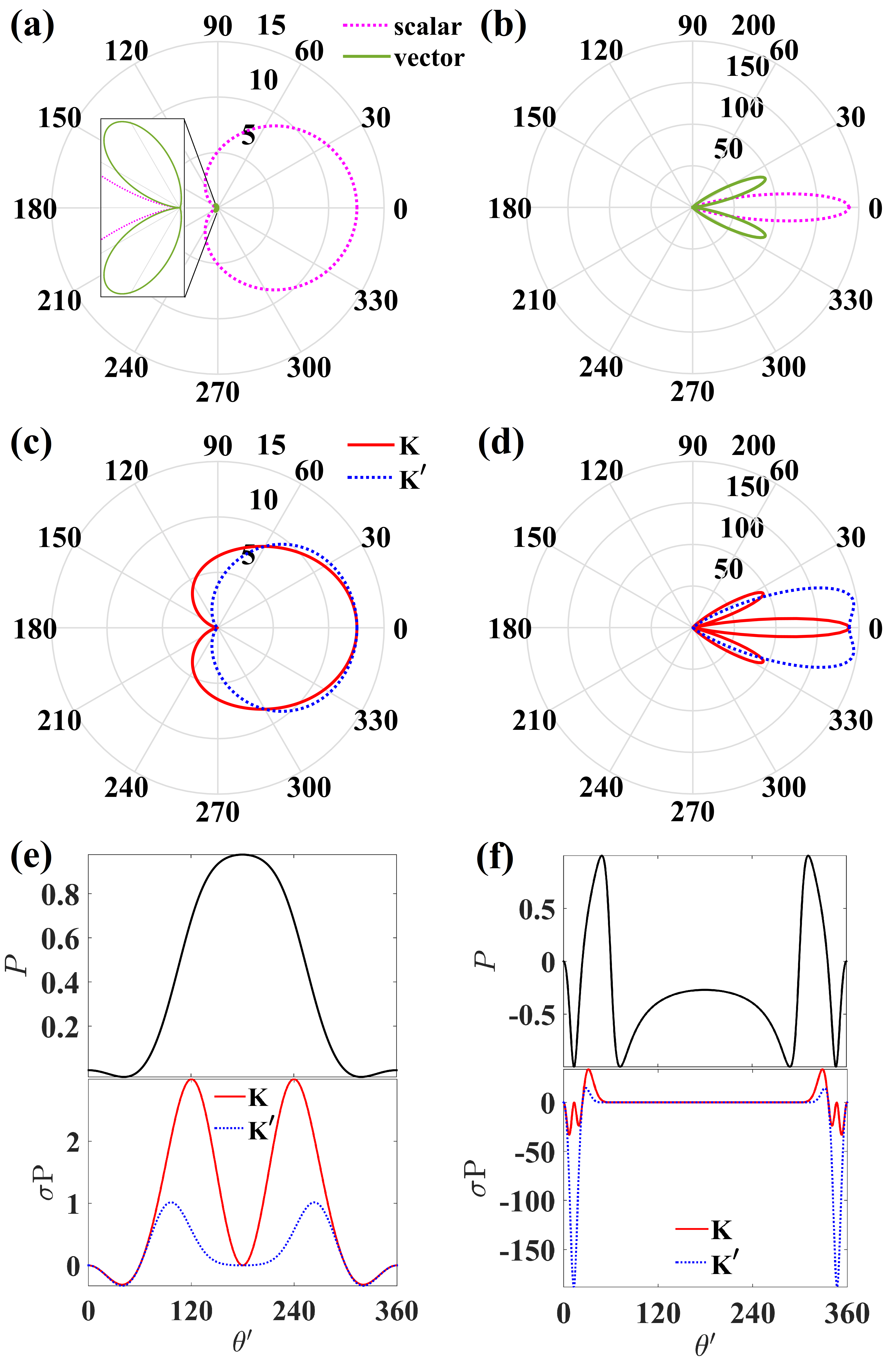}
	\caption{Polar plots of the differential cross section (\AA) for contributions of scalar (pink dotted) and pseudo-vector (green solid) potentials for valley $K$ and incident angle  $\theta'=0^\circ$. Panel (a) corresponds to energy $E=20$meV. Inset shows a zoom in to visualize the magnitude of the pseudo-vector potential contribution. Panel (b) corresponds to energy $E=300$meV. Panels (c) and (d): differential cross section per valley at low and high energy respectively for the total potential $V^{\tau} + \Phi$. Panels (e) and (f) show the corresponding polarization $P$ and $\sigma_DP$ for the two valleys. Other parameters: $b=15$nm, $\eta=0.1$, $g_s = 3$eV.}\label{cross_section_with_scalar}
\end{figure}

Results of similar scattering calculations but including the scalar potential within first order Born approximation are shown in Fig.~\ref{cross_section_with_scalar}. Panel (a), in the low energy regime, shows a much larger differential scattering cross section for the scalar field than for the pseudo-vector potential, while comparable values are obtained for the high energy regime as shown in panel (b). Panels (c) and (d) present results for the total scattering potential, i.e., $\Phi + V^{\tau}$ for valleys $K$ and $K'$ at low and high energy regimes respectively. In both cases, the structure still exhibits the valley filtering capability. However, the details of the valley polarization effect are dramatically different from those in the absence of the scalar potential as shown in panels (e) and (f).

As a final remark for this section, we want to emphasize that results presented are based on a perturbation expansion, namely, the Born approximation and its series expansion. For stronger potentials, i.e. larger values of $\eta$, however, quasi-bound states can exist. The Born series does not necessarily converge, and consequently scattering calculations are not reliable. For example, for a bump with $b=150$\AA\, and $\eta=0.2$, where $V_{max}\approx 2.3 E_b$, the 2nd order contribution becomes comparable to the 1st order. In this regime, the deformed region acts more like a quantum dot structure with quasi-bound states that act as resonant levels for transport. This is consistent with reported results\cite{Settnes}, where optimal valley polarization is predicted at resonant energies of a strong potential induced by a local Gaussian deformation. Resonant energies are strongly dependent on the details of the deformation and their determination needs to be done by appropriate modeling. The phenomena, that is also observed in fold-like deformations as we will show below, suggests that optimization of valley filtering properties in this regime requires a delicate tuning of the parameters of the deformation, in addition to precise angular resolution.

\section{Extended Gaussian fold deformation}
\label{SecIV}

In this section we analyze in detail the valley filtering properties of an extended deformation, i.e. a Gaussian fold with translation invariance along the $x$ (zigzag) direction (Fig.~\ref{deformations}(b)). In contrast to the previous case, we evaluate the transmission probabilities of electrons injected towards the fold using the numerical transmission matrix approach. We analyze the fold valley filtering properties as the incident angle for the current as well as its structural parameters are changed. We will first focus on the effect of the pseudo-vector potential. The effect due to the scalar potential will be discussed at the end of the section.

\subsection{Method for calculation of transmission probabilities}
We implement standard numerical transmission matrix methods to obtain transmission and reflection coefficients. The procedure involves real space discretization of either the pseudo vector potential $A_x(y)$ or the pseudomagnetic field $B_z(y)$, and matching wave functions in adjacent regions. Discretization of $B_z(y)$ avoids discontinuous changes in $A_x(y)$ that may produce spurious numerical effects. However, wave functions become non-trivial and the overall interpretation is less intuitive. We present results based on the discretization of $A_x(y)$ that are fully consistent with those obtained via $B_z(y)$ as discussed in Appendix~\ref{App_B}.

The gauge field $A_x(y)$ is split into $N-1$ slices in the interval $y\in[y_0,y_{N-1}]$, where the value for $A_{x,i}$ ($i=1,2\cdots N-1$) is chosen as the mid-point value of the continuum field in that region. We choose $y_0$ and $y_{N-1}$ symmetrically, i.e. $y_0=-y_{N-1}$, and large enough such that $A_x(y)$ is negligibly small outside this region.

To evaluate the transmission probability, the wave functions in different regions need to be calculated. Due to translation invariance along $x$ direction, the wave function in region $i$ reads $\Psi_i(x,y)=e^{ik_xx}\psi_i(y)$, with $k_x=k\cos\theta= \frac{E\cos\theta}{\hbar v_F}$, and $\theta\in[0,\pi]$ the incident angle. For positive energies ($E>0$), the wave function in the region $y<y_0$ is in general, a combination of spinors written as
\begin{equation}
\psi^\tau_0(y)=
\begin{pmatrix}
1\\e^{i\theta}
\end{pmatrix}
e^{ik_yy}+
r^\tau_0
\begin{pmatrix}
1\\e^{-i\theta}
\end{pmatrix}
e^{-ik_yy}
\label{incidwavefunction}
\end{equation}
where $\tau=\pm$ labels the valleys, $r_0^\tau$ is the reflection coefficient, and $k_y=k\sin\theta$.
In analogy, the wave function in region $y>y_{N-1}$ reads
\begin{equation}
\psi_N^\tau(y)=t_N^{\tau}
\begin{pmatrix}
1\\e^{i\theta}
\end{pmatrix}
e^{ik_yy}
\label{trasnmwavefunction}
\end{equation}
where $t_N^\tau$ is the transmission coefficient and we have assumed a current incident from the $y<y_0$ region.

Within the interval $[y_0,y_{N-1}]$, the wave function for valley $\tau$ in the $i$-th slice takes the form:
\begin{equation}
\begin{aligned}
\psi_i^\tau(y)&=t_i^\tau
\begin{pmatrix}
1\\\frac{\hbar v_F k_x-\tau v_F A_{x,i}+i\hbar v_Fq_{\tau,i}}{E}
\end{pmatrix}
e^{iq_{\tau,i}y}\\
&+ r_i^\tau
\begin{pmatrix}
1\\\frac{\hbar v_F k_x-\tau v_F A_{x,i}-i\hbar v_Fq_{\tau,i}}{E}
\end{pmatrix}
e^{-iq_{\tau,i}y}
\end{aligned}
\label{psi_in_fold}
\end{equation}
where 
\begin{equation}
q_{\tau,i}=\sqrt{k^2-\left(k_x-\tau \frac{A_{x,i}}{\hbar}\right)^2}
%\label{eqforq}
\label{q_in_fold}
\end{equation}
is the corresponding wave vector along the $y$ direction.

Wave functions in different slices are connected via the continuity condition. By employing the scattering matrix method\cite{ScatteringMatrixMethod}, one can ensure the continuity of the wave functions at the boundaries of each slice and solve for $r^\tau_0$ and $t_N^{\tau}$. The transmission probability can then be obtained from $T_\tau=\left|t_N^{\tau}\right|^2$.

For a non-zero transmission, we distinguish two different regimes: the scattering regime with $q_{\tau,i}$ real, i.e. propagating waves exist in the fold region; and the tunneling regime with $q_{\tau,i}$ imaginary, i.e, electrons tunnel through the fold. 
Let us focus first on the scattering regime. In this case Eq.~\ref{q_in_fold} imposes the constraint:
\begin{equation}
-k+\tau \frac{A_{x,i}}{\hbar}\le k_x \le k+\tau \frac{A_{x,i}}{\hbar}.
\end{equation}
At the same time, $k_x^2+k_y^2=k^2$ requires
\begin{equation}
-k \le k_x \le k.
\end{equation}
By combining these two conditions and considering that $A_{x,i}\le0$, we obtain
\begin{equation}
\begin{cases}
-k\le k_x \le k+\frac{A_{x,i}}{\hbar},&\text{$K$ valley}\\
-k-\frac{A_{x,i}}{\hbar} \le k_x \le k,&\text{$K'$ valley}
\end{cases}.
\end{equation}
As $k_x=k\cos\theta$, the above results indicate that the Gaussian fold allows different incident (transmission) windows for $K$ and $K'$ valleys:
\begin{equation}
\theta\in
\begin{cases}
\left[\arccos\left(1+\frac{v_FA_{x,min}}{E}\right),\pi\right],&\text{$K$ valley}\\
\left[0,\pi-\arccos\left(1+\frac{v_FA_{x,min}}{E}\right)\right],&\text{$K'$ valley}
\end{cases}
\label{T_window},
\end{equation}
where $A_{x,min}$ represents the minimum value of $A_x$.
Therefore, valley polarization is expected in the transmitted beams if the incident angle is chosen appropriately. Furthermore, when $\pi-\arccos\left(1+\frac{v_FA_{x,min}}{E}\right)\le\arccos\left(1+\frac{v_FA_{x,min}}{E}\right)$, i.e. $E\le-v_FA_{x,min}$, the transmission spectra of $K$ and $K'$ valleys are completely separated. 

The tunneling regime, with $q_{\tau,i}$ imaginary is solved numerically. Results for both regimes are discussed below.

\subsection{Results and discussion}

\begin{figure}[ht]
\includegraphics[width=3.4in]{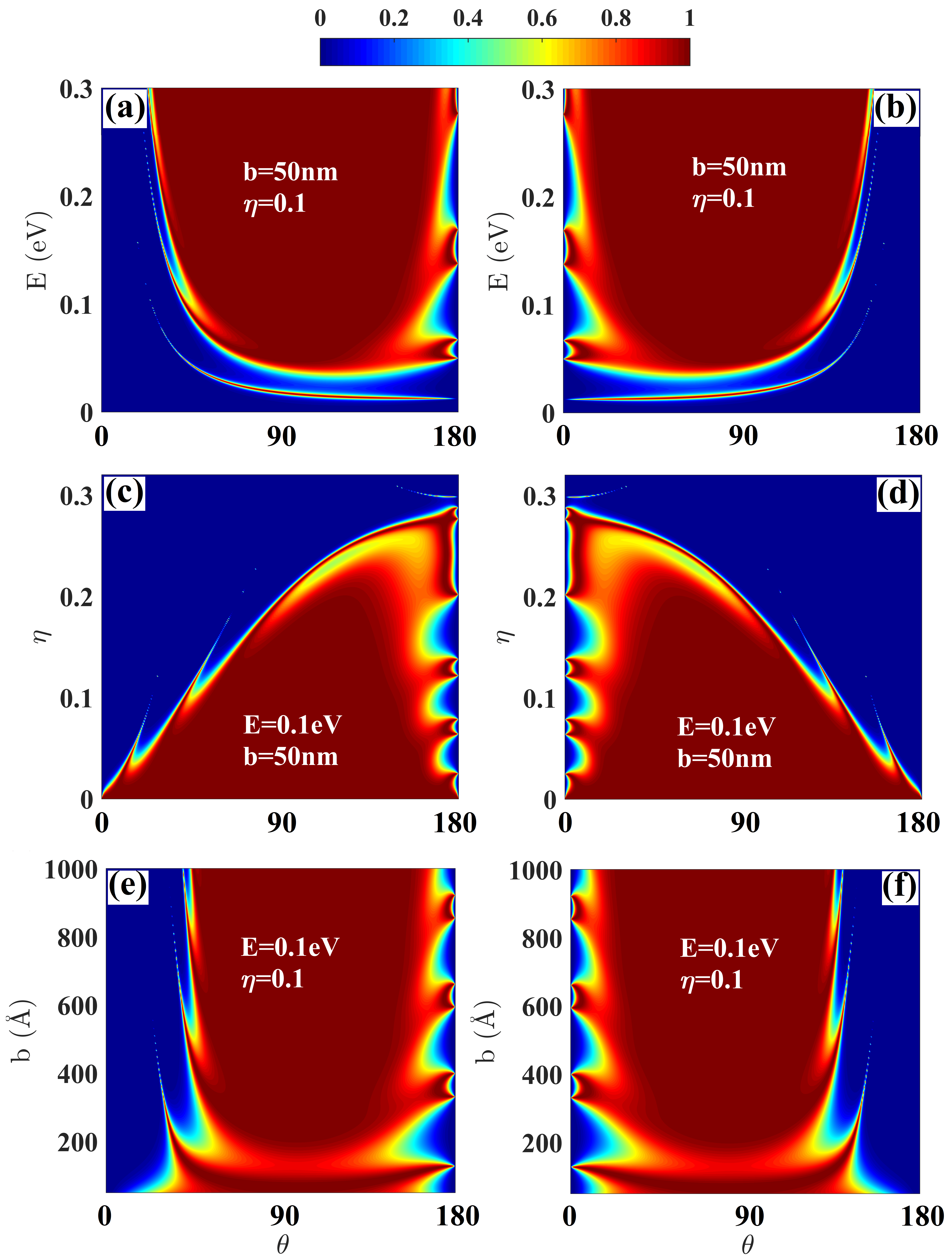}
\caption{Transmission probability versus incident angle $\theta$ for $K$ (left column) and $K'$ (right column)  valleys.  Results are shown as a function of energy $E$ in panels (a)-(b), as a function of the strain strength $\eta$ in panels (c) and (d), and as a function of the fold width $b$ in panels (e) and (f) for each valley respectively.}
\label{Transmission_K_vs_Kprime}
\end{figure}

Fig.~\ref{Transmission_K_vs_Kprime} shows numerical results for the transmission probability $T_K$ and $T_K'$ for valleys $K$ and $K'$ respectively, as functions of the incident angle $\theta$ (normalized by $\pi$) for different values of energy (panels (a) and (b)), strain strength (panels (c) and (d)) and fold width $b$ (panels (e) and (f)). Clearly, valley polarization is observed, i.e. $T_K(\theta)\ne T_{K'}(\theta)$, for a wide range of values of $\theta$. Regions with high transmission probability (dark red) are consistent with Eq.~\ref{T_window} (see also Fig.~\ref{T_vs_E_Ex} in Appendix~\ref{App_B}). Note that transmission probabilities satisfy $T_K(\theta)=T_{K'}(\pi-\theta)$, consistent with the fact that the pseudo-vector potential respects time-reversal symmetry.
In addition to the solid color regions, sharp lines in panels (a) and (b) with finite transmission are distinguishable. These correspond to energies in the 'tunneling regime', i.e., with imaginary values of $q_{\tau,i}$.

The figure shows that transmission probability and valley polarization effects are robust for a large range of energies (in addition to the isolated resonant ones) and can be tuned via the geometric parameters of the deformation, i.e. $\eta$ and $b$ for any given energy.

\begin{figure}[ht]
\includegraphics[width=3.4in]{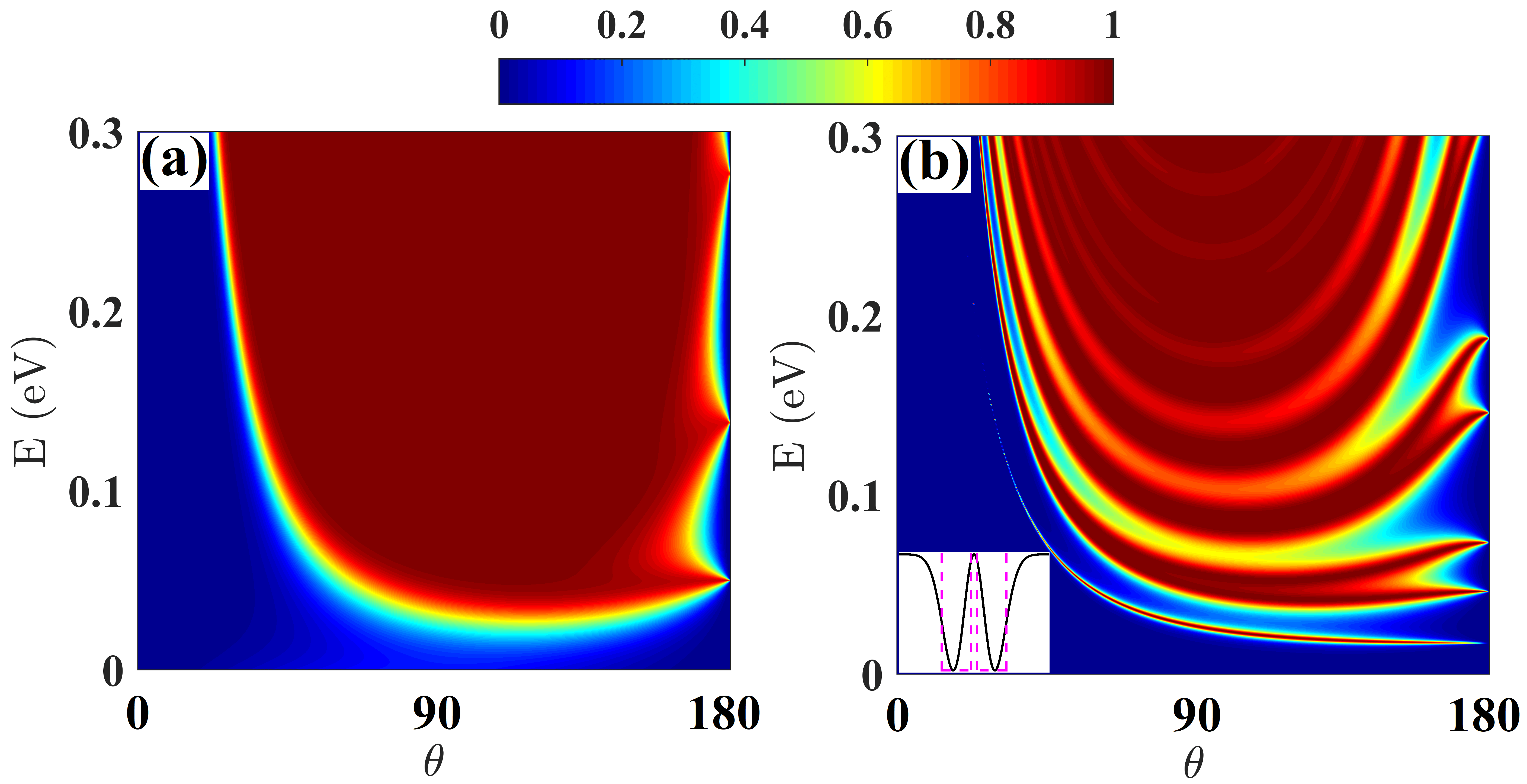}
\caption{Transmission probability versus incident angle $\theta$ for $K$ valley. Results are shown as a function of energy $E$. Panel (a): transmission for a single barrier potential. Panel (b): transmission for a double square barrier potential. Inset: solid (black) line: profile of $\bA(y)$, dashed (magenta) line: profile of double square potential well. Parameters: $b=50$nm, $\eta=0.1$.}
\label{Double_barrier}
\end{figure}

The profile of the transmission spectra can be intuitively understood by considering a simple model of a symmetric double square pseudo-vector potential well structure as shown in the inset of panel (b) of Fig.~\ref{Double_barrier} in dashed (magenta) lines, superimposed to the real pseudo-vector potential profile in solid (black) lines. 

The transmission spectrum of the symmetric double square pseudo-vector potential well appears more definite due to the sharp structure of the edges, however it reproduces the main features shown in Fig.~\ref{Transmission_K_vs_Kprime}(a), especially resonances at $\theta=\pi$ and nearby. These resonances, already present in a single well potential (corresponding to one of the wells in $A_x(y)$) spectrum as shown in Fig.~\ref{Double_barrier} (a), split due to the two-well structure. In the case of a symmetric double square pseudo-vector potential well model, the resonances due to a single well occur at $qd=n\pi$, where $q$ is the momentum inside the well, $d$ is the width of the well, and $n$ is an integer. The solution of a double well model predicts the position of their splittings and widths\cite{DoubleWell}.

\begin{figure}[ht]
\includegraphics[width=3.4in]{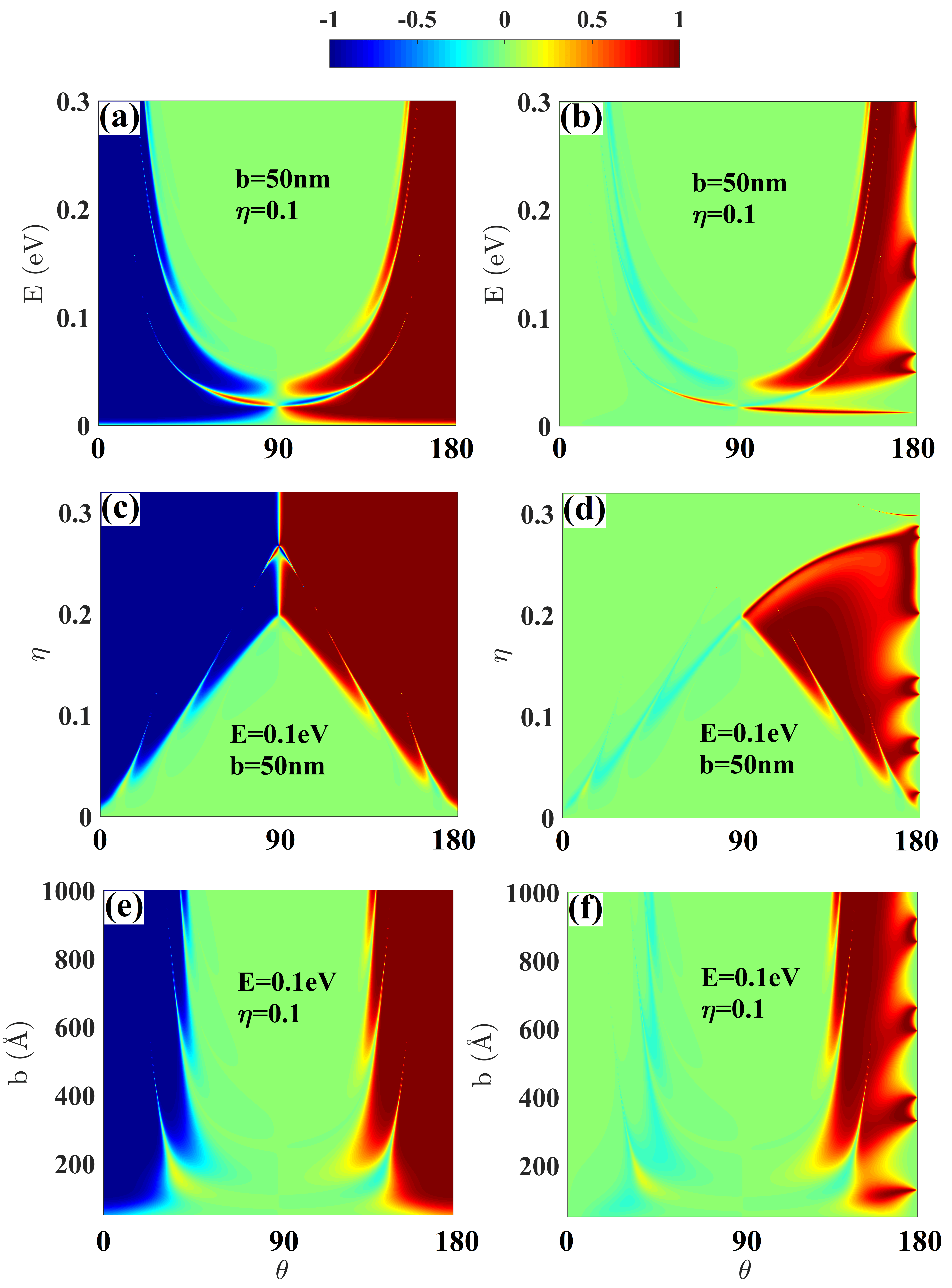}
\caption{Valley polarization $P$ (left column) and $T_KP$ (right column) versus incident angle $\theta$ and energy $E$, strain strength $\eta$, and fold width $b$.}
\label{P_and_TP}
\end{figure}

In order to characterize the efficiency of the Gaussian fold for inducing valley polarized transmitted currents, we redefine the angle dependent polarization coefficient in terms of transmission coefficients as:
\begin{equation}
P=\frac{T_K-T_{K'}}{T_K+T_{K'}}
\end{equation}
In the left column of Fig.~\ref{P_and_TP} we present results for  $P$ corresponding to data shown in Fig.~\ref{Transmission_K_vs_Kprime} for both valleys. $P$ exhibits a mirror symmetric structure with respect to $\theta=\pi/2$ as expected.
Large polarization regions appear at large and small incident angles with respect to the fold axis (oriented along the zigzag direction) at a large range of parameters. 

In analogy with the analysis carried out for the bump deformation, we note that the degree of polarization is not enough to ensure a measurable detection of polarized currents. In addition to high polarization values, a detectable signal must  have a large transmission probability.  Thus, it is convenient to evaluate the product of these two quantities as an indicator of the efficacy of the fold as a valley polarizer. The right column in Fig.~\ref{P_and_TP}, shows the product of the transmission probability $T$ and valley polarization $P$ for $K$ valley.
 
A large parameter region (red area in Fig.~\ref{P_and_TP} right column) with large values of both $T$ and $P$ can be identified, indicating high efficiency for a wide range of energies. $TP$ results for valley $K'$ are obtained from this data by applying the relation $T_{K'}(\pi-\theta)P(\pi-\theta)=-T_{K}(\theta)P(\theta)$ which results on  an antisymmetric profile with respect to $\theta=\pi/2$ (not shown here).

In addition to the aspects considered above, the fold axis orientation affects the polarization and magnitude of transmitted currents. The influence of the orientation can be seen by considering an arbitrary direction $\gamma$ for the fold axis with respect to the zigzag direction. By choosing the $\hat{x}$ direction along the fold axis, the pseudo-magnetic field in this case is obtained by applying the appropriate rotation, and transforms as $\textbf{B}\rightarrow \textbf{B}\cos(3\gamma)$\cite{RotationEffectStrainPRB2015,RotationEffectStrainPRB2016}. Results from Eq.~\ref{T_window} remain valid provided the transformation $A_x\rightarrow A_x\cos(3\gamma)$ is performed (see details in Appendix~\ref{App_C}). 

Fig.~\ref{Fold_along_25_deg} shows the results for a fold along $\gamma=25^\circ$ respective to the zigzag direction. The large transmission region increases (Red area in Fig.~\ref{Fold_along_25_deg}(a,b)) due to a smaller pseudo-vector potential barrier ($A_x\rightarrow A_x\cos(3\gamma)\approx0.26A_x$), and as a consequence, the extent of the valley polarized regime decreases (Fig.~\ref{Fold_along_25_deg} (c)). It is clear, from the nature of the rotation that folds along the zigzag (armchair) direction, e.g. $\gamma=0^\circ\,(30^\circ)$, will yield the largest (smallest) valley polarization effect.

\begin{figure}[ht]
\includegraphics[width=3.4in]{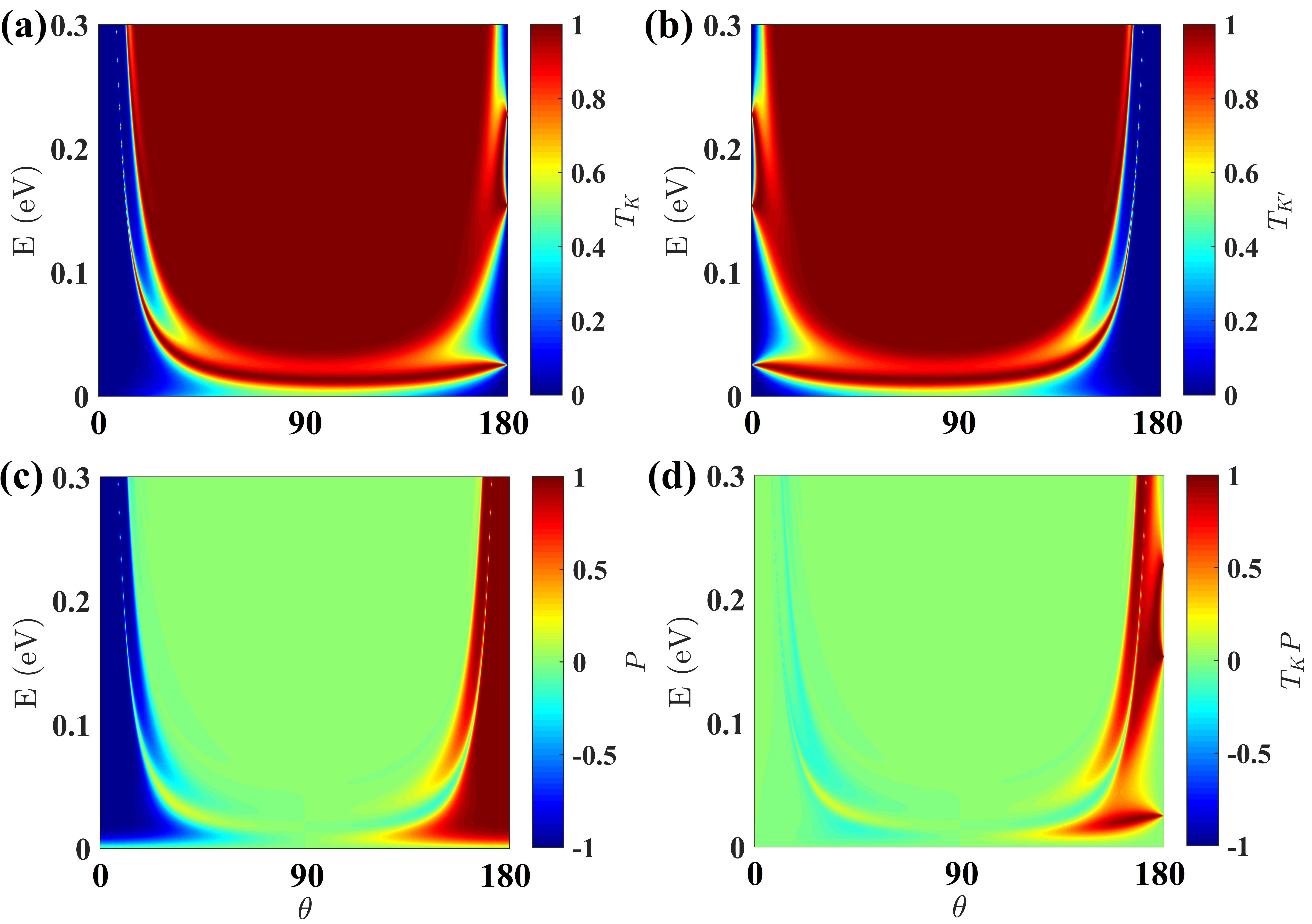}
\caption{Results for Gaussian fold with axis oriented at an angle $\gamma=25^\circ$ with respect to the zigzag direction. (a, b) Transmission $T$ for $K$ and $K'$ valley, (c) Polarization $P$, and (d) $TP$ for $K$ valley. Other parameters: $b=50$nm, $\eta=0.1$.}
\label{Fold_along_25_deg}
\end{figure}

\begin{figure}[ht]
\includegraphics[width=3.4in]{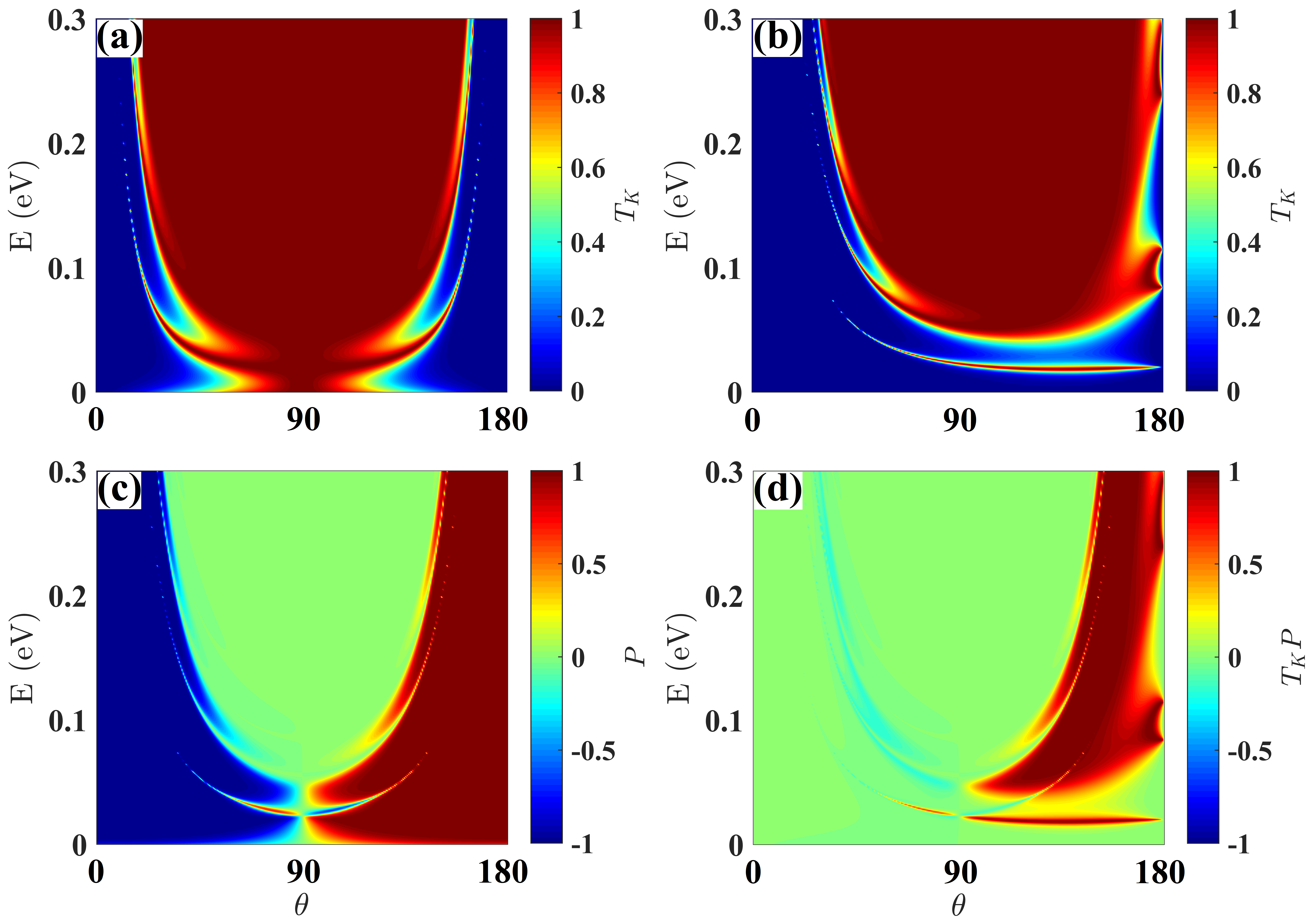}
\caption{Transmission $T$ due to (a) only the scalar potential and (b) both scalar and vector potentials for $K$ valley. The scalar potential contribution is valley independent, while (b) is mirror symmetric with respect to $\theta=\pi/2$ for $K'$ valley. (c) Valley polarization $P$, and (d) $TP$ for $K$ valley.}
\label{Fold_scalar_effect}
\end{figure}

Now let's discuss the effect of the scalar potential. The above formalism remains valid as long as the replacements $E\rightarrow E-\Phi_i$ and $k\rightarrow \left| k-\Phi_i/(\hbar v_F)\right|$ are performed in Eqs.~\ref{psi_in_fold},~\ref{q_in_fold} and the corresponding following discussions. One can easily see that the effect of the scalar potential is to shift the energy, as clearly shown in Fig.~\ref{Fold_scalar_effect}(b) where the transmission window is shifted upward in energy. Due to the spatial dependence of the scalar potential and its nontrivial effect on transmission coefficients as shown in  Fig.~\ref{Fold_scalar_effect}(a), the profile of the transmission spectrum is slightly different from that without the scalar potential in Fig.\ref{Transmission_K_vs_Kprime}(a). However, notice that the valley polarization effect is preserved, and large polarization and strong transmission coexist as shown in Fig.~\ref{Fold_scalar_effect}(c,d).

\section{Discussion of proposed experimental setups}
\label{SecV}

In Sec.~\ref{SecIII} we showed that a Gaussian bump can induce valley polarized currents, where the magnitude of polarization can be controlled via appropriate tuning of geometrical parameters that determine the strain (the ratio $\eta = h_0/b$), the incident energy $E$ or the width of the bump $b$. Furthermore, the sign of the polarization can be reversed via rotating the incident angle by $60^\circ$. However, generation of simultaneous large scattering cross section and high polarization values appears quite challenging due to the narrow angular regions and overall small size of transmitted signals. In the regime beyond perturbation theory, where the bump is represented by a large potential, it is possible to obtain large transmission signals at resonant energies\cite{Settnes}. This regime however limits the tunability of the device and requires rather precise determination of resonant energies.

Fortunately, valley filtering properties of folds appear more promising. As shown above, incident currents at a wide variety of angles show good transmission and polarization properties, that can be tuned by modifying the geometrical parameters in a large range of values. Furthermore, folds with different orientations with respect to crystalline axis, remain good valley filters, with those aligned along zigzag directions being the most efficient.

In all these results, samples were considered to be pristine, i.e., disorder introduced either by impurities or local crystalline defects was neglected. These disorder sources, being local in nature, enhance inter-valley scattering and lead to a reduced polarization effect. However, for the strain-induced mechanism proposed above, valley filtering refers to the {\it spatial} separation of states originated in different valleys. The range of the spatial separation is determined by the effective width of the deformation for folds and angular resolution for bumps. As these are tuning parameters, it is conceivable that the role of disorder may be made inconsequential, specially for rather clean samples as the ones envisioned in these proposals.

\begin{figure}[ht]
\centering
\includegraphics[width=3.4in]{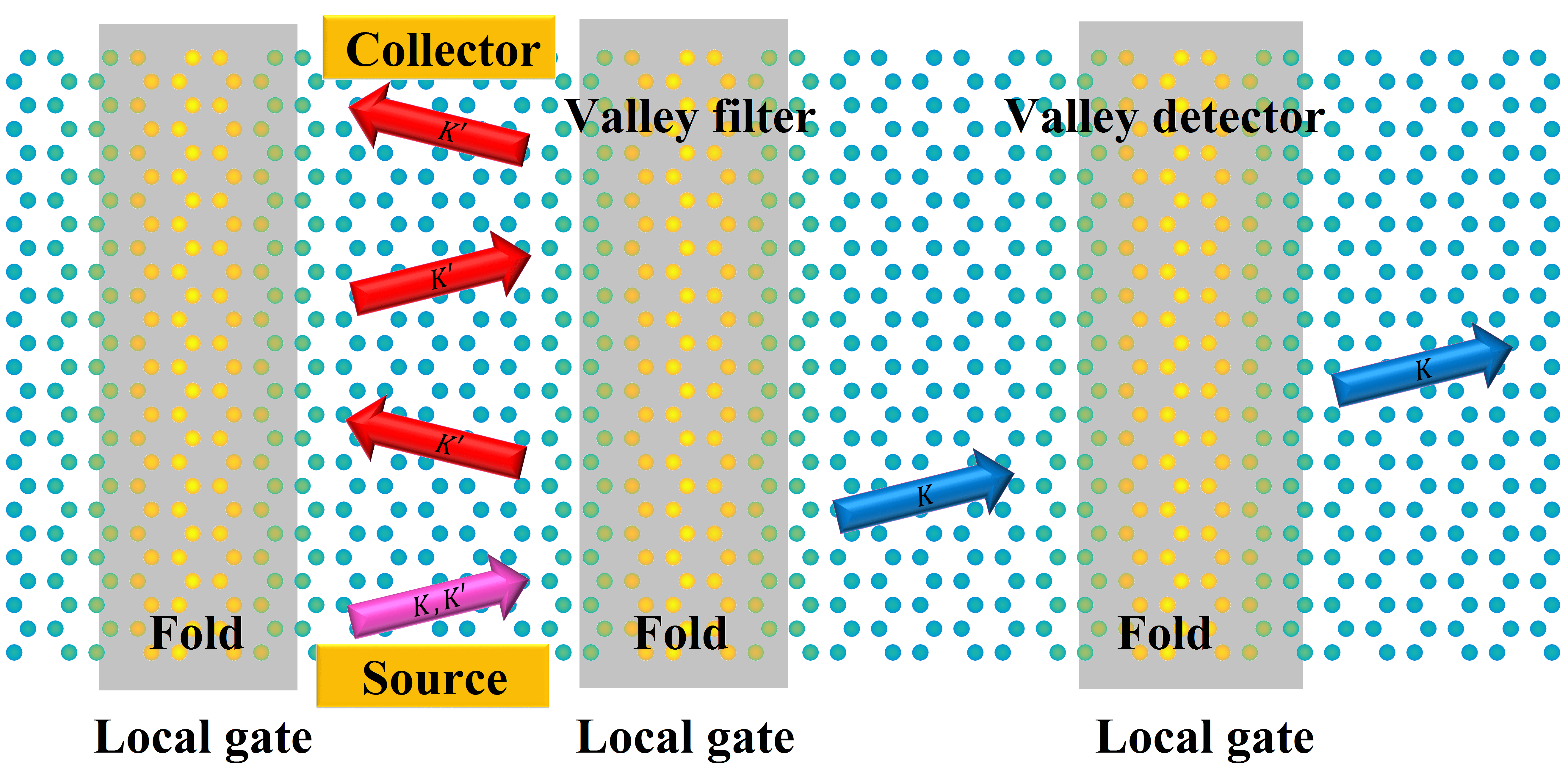}
\caption{Scheme of valley filtering and detection of polarized beams with Gaussian folds (yellow areas). The local gates (grey areas) are employed to tune the Fermi levels so that the dependence of transmission probability on incident energy can be measured. In the middle trench setup (see text), multiple reflection and transmission events between the two folds can generate highly polarized electron beams at the end of the trench.}
\label{expt_setup}
\end{figure}

As for the generation and detection of fully valley polarized currents, proposed setups would employ a double fold structure. Fig.~\ref{expt_setup} shows two different setups. In one, the first fold (middle fold in the figure) acts as valley polarizer to generate valley polarized transmitted beams while a set of secondary folds (side folds in the figure) would act as detectors. Alternatively, a double-fold structure may be used to collect highly polarized electron beams in the trench formed between two folds (left and center folds in Fig.~\ref{expt_setup}). This does not require a perfect valley filtering by the fold (i.e. $T_K=1$) since multiple reflection and transmission events in the trench will make the remaining electrons highly $K'$ polarized, akin to a Fabry-Perot interferometer.

{\it Note added:} While preparing this manuscript we became aware of the work by R. Carrillo-Bastos et al.\cite{RamonBump2018}, T. Stegmann et al. \cite{SzpakBump2018}, and E. Mu\~noz et al.\cite{EnriqueBubble2017}, where similar graphene bubble systems were studied and valley polarization phenomena were observed.

\begin{acknowledgments}
We acknowledge support from NSF-DMR 1508325. Portions of this work were completed at the Aspen Center for Physics under support from NSF-PHY-1066293.
\end{acknowledgments}

\appendix

\section{Evaluation of $V^{\tau}_{\bk,\bk'}$, $(VGV)^{\tau}_{\bk,\bk'}$, and $\Phi_{\bk,\bk'}$}
\label{App_A}

We evaluate $(VGV)^{\tau}_{\bk,\bk'}$ which requires the expression for $V^{\tau}_{\bk,\bk'}$, the Fourier transform of $V^{\tau}(\br)$. Using:

\begin{equation}
G(\br',\br'')=\frac{1}{(2\pi)^2}\int e^{i\textbf{p}\cdot(\br'-\br'')}G(\textbf{p},E)d\textbf{p},
\end{equation}
where 
\begin{equation}
\begin{aligned}
E&=\hbar v_F k\\
G(\textbf{p},E)&=\left[E-H_0(\textbf{p})\right]^{-1}\\
&=\frac{E+H_0(\textbf{p})}{(E+i0^+)^2-\hbar^2v_F^2p^2}\\
H_0(\textbf{p})&=\hbar v_F p
\begin{pmatrix}
0&e^{-i\theta_p}\\
e^{i\theta_p}&0
\end{pmatrix}\\
\theta_p&=\angle\textbf{p}
\end{aligned},
\end{equation}
where $\theta_p$ is measured with respect to the zigzag crystalline orientation. 

From the definition of $(VGV)^{\tau}_{\bk,\bk'}$:
\begin{equation}
(VGV)^{\tau}_{\bk,\bk'}=\braket{\psi_{\bk}(\br')|V^{\tau}(\br')G(\br',\br'')V^{\tau}(\br'')|\psi_{\bk'}(\br'')}
\end{equation}
one can write 
\begin{equation}
\begin{aligned}
&(VGV)^{\tau}_{\bk,\bk'}\\
&=\int \frac{d\textbf{p}}{(2\pi)^2}\braket{u_{\bk}|\tilde{V}^\tau(\bk-\textbf{p})G(\textbf{p},E)\tilde{V}^\tau(\textbf{p}-\bk')|u_{\bk'}}
\end{aligned},
\end{equation}
with
\begin{equation}
\tilde{V}^\tau(\textbf{q})=\int e^{-i\textbf{q}\cdot\br}V^\tau(\br)d\br
\end{equation}
the Fourier transform of the scattering potential.

To evaluate $\tilde{V}^\tau(\bk-\textbf{p})$ explicitly notice that
\begin{equation}
\begin{aligned}
e^{-i(\bk-\textbf{p})\cdot\br}&=e^{i(c\cos\phi+d\sin\phi)}\\
&=e^{iR(\bk,\textbf{p})\sin\left[\phi+\alpha(\bk,\textbf{p})\right]}
\end{aligned}
\end{equation}
where
\begin{equation}
\begin{aligned}
c&=-kr\cos\theta+pr\cos\theta_p\\
d&=-kr\sin\theta+pr\sin\theta_p\\
\alpha(\bk,\textbf{p})&=\arctan\left(\frac{c}{d}\right)=\arctan\left(\frac{-\cos\theta+\tilde{p}\cos\theta_p}{-\sin\theta+\tilde{p}\sin\theta_p}\right)\\
R(\bk,\textbf{p},r)&=\sqrt{c^2+d^2}=kr\sqrt{1+\tilde{p}^2-2\tilde{p}\cos(\theta-\theta_p)}\\
\tilde{p}&=\frac{p}{k}
\end{aligned}.
\end{equation}

Furthermore, in order to evaluate the integral of the Fourier transform, we will employ
\begin{equation}\label{angular_integral_formula}
J_n(z)=\frac{1}{2\pi}\int_{-\pi}^{\pi}e^{-in\theta}e^{iz\sin\theta}d\theta,
\end{equation}
and
\begin{equation}\label{radial_integral_formula}
\int_{0}^{\infty}J_\mu(at)e^{-\gamma^2t^2}t^{\mu+1}dt=a^{\mu}(2\gamma^2)^{-\mu-1}e^{-\frac{a^2}{4\gamma^2}}
\end{equation}
when $\Re\left(\mu\right)>-1$ and $\Re\left(\gamma^2\right)>0$.

One can verify that the angular integral yields
\begin{equation}
\begin{aligned}
\int_{0}^{2\pi}e^{iR\sin(\phi+\alpha)}e^{i2\phi}d\phi&
=e^{-i2\alpha}\int_{0}^{2\pi}e^{iR\sin\phi}e^{i2\phi}d\phi\\
&=2\pi e^{-i2\alpha}J_2(R)\\
\int_{0}^{2\pi}e^{iR\sin(\phi+\alpha)}e^{-i2\phi}d\phi&
=e^{i2\alpha}\int_{0}^{2\pi}e^{iR\sin\phi}e^{-i2\phi}d\phi\\
&=2\pi e^{i2\alpha}J_2(R)
\end{aligned}
\end{equation}

The radial integral gives
\begin{equation}
\int_{0}^{\infty}g\left(\frac{r}{b}\right)J_2(R)rdr=\frac{b^2}{4}g\left(\frac{1}{4}R(\bk,\textbf{p},b)\right).
\end{equation}

Therefore,
\begin{equation}
\tilde{V}^\tau(\bk-\textbf{p})=-\tau\frac{\pi}{2}\overline{\beta}\eta^2b^2g_1
\begin{pmatrix}
0&e^{-i2\alpha_1}\\
e^{i2\alpha_1}&0
\end{pmatrix},
\end{equation}
where 
\begin{equation}
\begin{aligned}
g_1&=g\left(\frac{1}{4}R(\bk,\textbf{p},b)\right)\\
\alpha_1&=\alpha(\bk,\textbf{p})
\end{aligned}.
\end{equation}

In the case of $\textbf{p}=\bk'$, i.e. the incident wave vector, we have $\alpha(\bk,\bk')=-\theta_+$ and $R(\bk,\bk',b)=2kb|\sin\theta_-|$, where $\theta_-=(\theta-\theta')/2$ and $\theta_+=(\theta+\theta')/2$. Using these results we obtain
\begin{equation}\label{vector_potential_matrix_element}
\begin{aligned}
V^{+}_{\bk,\bk'}&=-V^{-}_{\bk,\bk'}\\
&=-\frac{1}{2}\pi\overline{\beta}b^2\eta^2\cos(3\theta_+)g\left(\frac{1}{2}kb\sin\theta_-\right)
\end{aligned}
\end{equation}
which is a real function.

By the same token, one can obtain 
\begin{equation}
\tilde{V}^\tau(\textbf{p}-\bk')=-\tau\frac{\pi}{2}\overline{\beta}\eta^2b^2g_2
\begin{pmatrix}
0&e^{-i2\alpha_2}\\
e^{i2\alpha_2}&0
\end{pmatrix},
\end{equation}
where 
\begin{equation}
\begin{aligned}
g_2&=g\left(\frac{1}{4}R(\bk',\textbf{p},b)\right)\\
\alpha_2&=\alpha(\bk',\textbf{p})
\end{aligned}.
\end{equation}

Combing all the above results it is straightforward to get
\begin{equation}
\begin{aligned}
&(VGV)^{\tau}_{\bk,\bk'}=\left(\frac{kb}{4}\overline{\beta}\eta^2b\right)^2\iint d\theta_p d\tilde{p}\frac{Eg_1g_2\tilde{p}}{(E+i0^+)^2-E^2\tilde{p}^2}\\
&\times\left[\cos\left(\theta_--2\alpha_1+2\alpha_2\right)+\tilde{p}\cos\left(\theta_+-2\alpha_1-2\alpha_2+\theta_p\right)\right]
\end{aligned}
\end{equation}
which is valley-independent.

To evaluate $\Phi_{\bk,\bk'}$, one can easily check that
\begin{equation}
\begin{aligned}
\Phi_{\bk,\bk'}&=g_s\eta^2\int e^{-i(\bk-\bk')\cdot\br}g\left(\frac{r}{b}\right)\cos\theta_-d\br\\
&=g_s\eta^2\cos\theta_-\iint e^{i2kr\sin(\theta_+-\phi)\sin\theta_-}g\left(\frac{r}{b}\right)rdrd\phi
\end{aligned}
\end{equation}
where $\theta_-=(\theta-\theta')/2$ and $\theta_+=(\theta+\theta')/2$. From Eq.~\ref{angular_integral_formula}, the integral over $\phi$ will yield a Bessel function such that
\begin{equation}
\Phi_{\bk,\bk'}=2\pi g_s\eta^2\cos\theta_-\int J_0(2kr\sin\theta_-)g\left(\frac{r}{b}\right)rdr.
\end{equation}
Using integration by parts and Eq.~\ref{radial_integral_formula} one can solve the integral over $r$, which is given by
\begin{equation}
\Phi_{\bk,\bk'}=\frac{1}{2}\pi b^2g_s\eta^2\cos\theta_-\left[1-\frac{1}{2}(kb\sin\theta_-)^2\right]e^{-\frac{1}{2}(kb\sin\theta_-)^2}.
\end{equation}
By comparing this equation with Eq.~\ref{vector_potential_matrix_element} one can easily realize that, in the low energy regime ($kb\ll 1$), scattering due to the scalar potential is much stronger than its pseudo-vector potential counterpart, as shown by the numerical results presented in the main text.

\section{Evaluation of the transmission probability by discretization of the pseudo-magnetic field}
\label{App_B}

\begin{figure}[ht]
\includegraphics[width=3.4in]{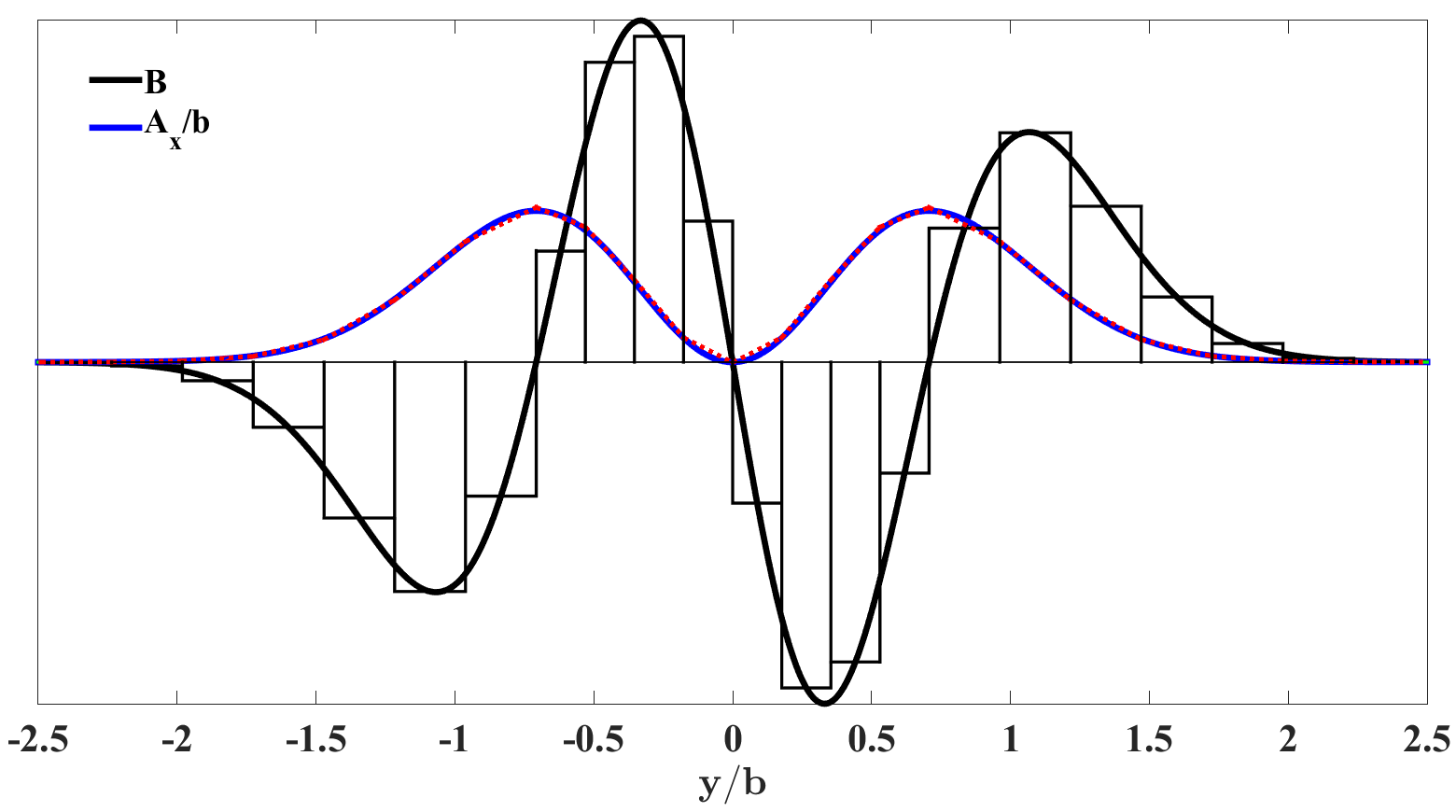}
\caption{Scheme for discretization of the pseudo-magnetic field into constant slices (black rectangles) and corresponding gauge field segments (red dashed lines). Thick black and blue curves represent continuous pseudo-magnetic field $B_z$ and the gauge fields $A_x/b$, respectively.}
\label{Const_B_split}
\end{figure}

In an analogous procedure to the one described in the main text, the pseudo-magnetic field $B_z$ is split into $N-1$ slices in the region $y\in[y_0,y_{N-1}]$, where in each slice $B_{z,i}$ ($i=1,2\cdots N-1$) is taken as a constant (black lines in Fig.~\ref{Const_B_split}) equal to the mid-point value. We choose $y_0$ and $y_{N-1}$ symmetrically, i.e. $y_0=-y_{N-1}$, and assume that they are large enough such that $B_z$ can be considered zero outside this region. As $B_z=-\frac{dA_x}{dy}$, in each slice we have $A_{x,i}(y)=C_i-B_{z,i}(y-y_{i-1})$, where $C_1=0$ and $C_i=-\sum_{j=1}^{i-1}B_{z,j}(y_j-y_{j-1})$ for $i>1$ (red dashed lines in Fig.~\ref{Const_B_split}).

Due to translation invariance, the wave function in region $i$ reads $\Psi_i(x,y)=e^{ik_xx}\psi_i(y)$, where $k_x$ is the momentum in $x$ direction. For the field-free regions $y<y_0$ and $y>y_{N-1}$, the wave functions are those given in Eqs.~\ref{incidwavefunction}, \ref{trasnmwavefunction}. Inside the region $[y_0,y_{N-1}]$, the wave functions are non-trivial to obtain, and special techniques are required to obtain those at small values of $B_{z}$. For constant magnetic field, the wave functions are commonly expressed in terms of parabolic cylinder functions\cite{ParabolicCylinderPRL,ParabolicCylinderPRB}. However, these wave functions diverge when the magnetic field is vanishing. In the present case, because the magnetic field vanishes at three points (thick black curve in Fig.\ref{Const_B_split}), special care is needed for using parabolic functions. Instead, we employ the series method and propose well-behaved solutions in all slices in the whole region $[y_0,y_{N-1}]$\cite{SeriesSolution}. 

Using the scattering matrix method\cite{ScatteringMatrixMethod} to ensure the continuity of the wave functions at the boundaries of each slice, one can solve for $r^\tau_0$ and $t_N^{\tau}$. The transmission probability is then obtained from $T_\tau=\left|t_N^{\tau}\right|^2$, which is consistent with that described in the main text.

\begin{figure}[ht]
\includegraphics[width=3.4in]{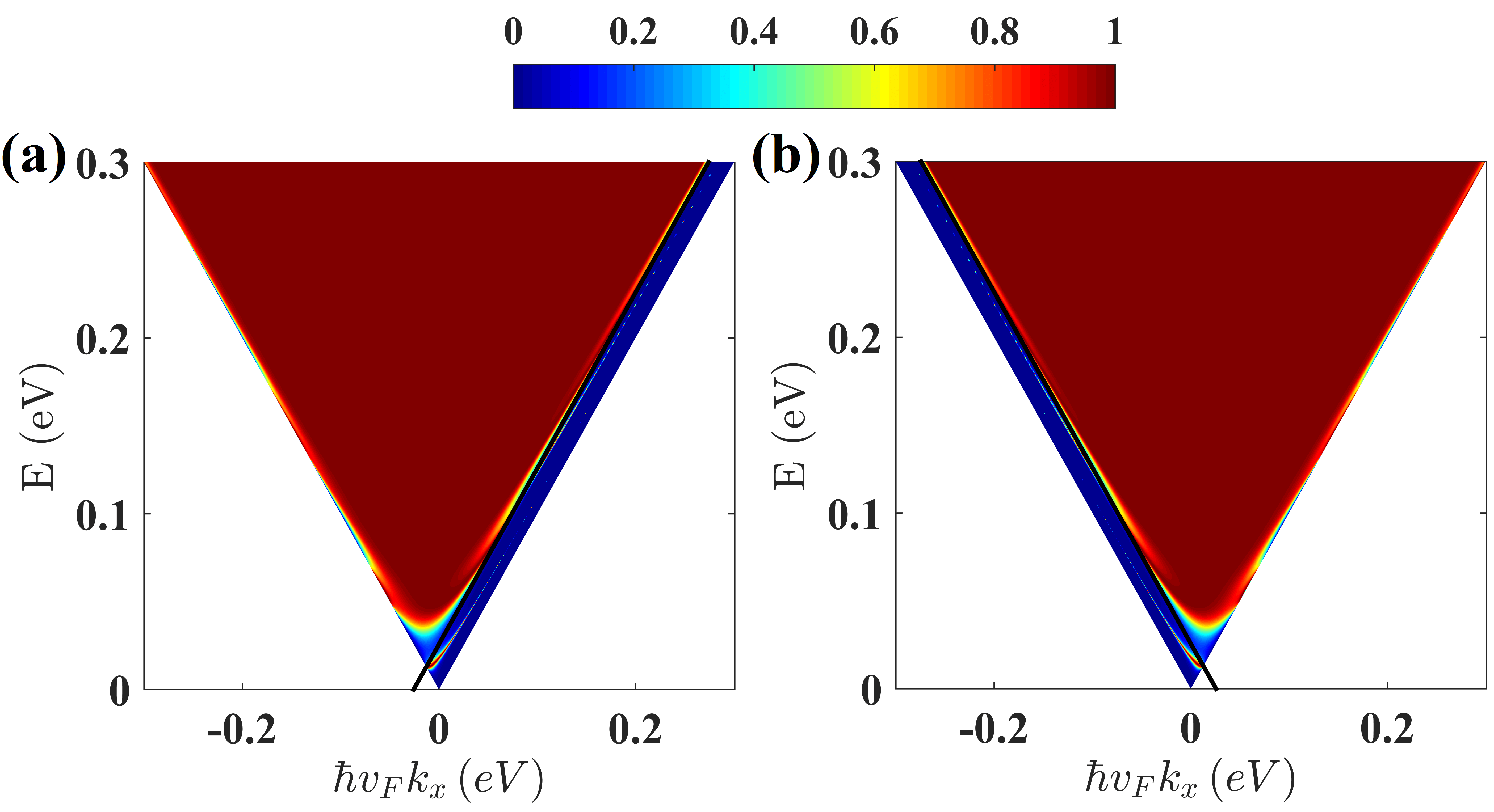}
\caption{Transmission probability versus $E\cos\theta$ and energy $E$ for (A) $K$ and (B) $K'$ valley. Parameters: $b=50$nm, $\eta=0.1$. The black lines correspond to the left ($K$) and right ($K'$) boundaries of Eq.~\ref{T_window}.}
\label{T_vs_E_Ex}
\end{figure}

\section{Graphene with a Gaussian fold along an arbitrary direction}
\label{App_C}

\begin{figure}[ht]
\centering
\includegraphics[width=3in]{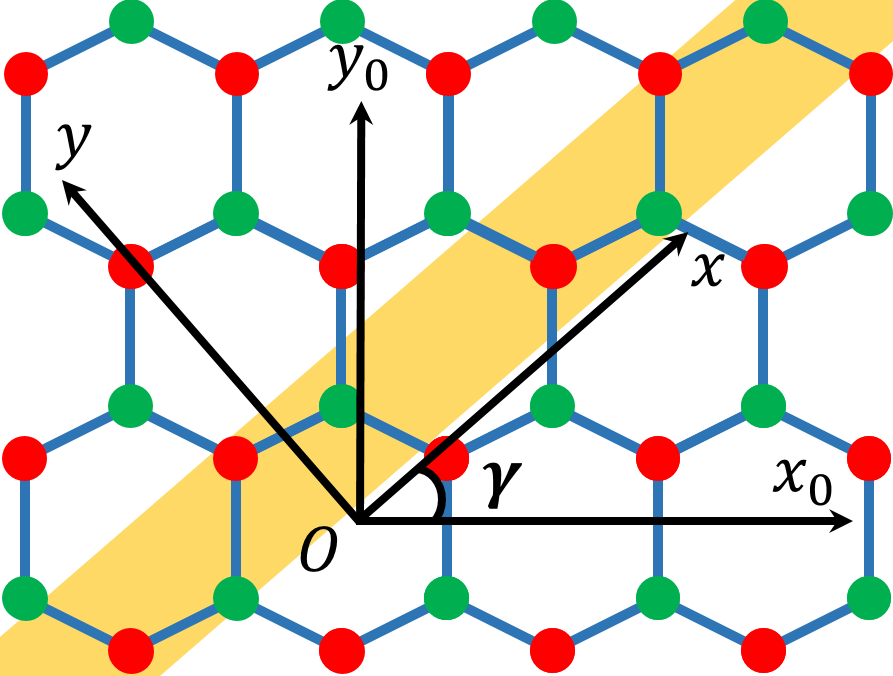}
\caption{A Gaussian fold (yellow shaded area) along angle $\gamma$ with respect to the zigzag direction. The zigzag and armchair directions define the original $x_0oy_0$ reference frame. The rotated frame $xoy$ is defined with the $x$ axis lying along the fold axis.}
\label{Rotate_Fold}
\end{figure}

Fig.\ref{Rotate_Fold} shows the schematics of a Gaussian fold along an arbitrary direction $\gamma$ with respect to the zigzag crystalline orientation. For simplicity, we define a rotated frame $xoy$ with the $x$ axis along the axis of the fold. For the case of $\gamma=0$, the coordinate frame is named $x_0oy_0$, and all the quantities in this frame will be identified by an index $0$ in the following discussion.

For $\gamma=0$, the gauge field is given by
\begin{equation}
\textbf{A}^0=\left(A^0_x,A^0_y\right)^T=\frac{\hbar\beta}{2a}\left(\epsilon^0_{xx}-\epsilon^0_{yy},-2\epsilon^0_{xy}\right)^T.
\end{equation}

For $\gamma\ne0$, the (counter-clockwise) rotated frame is obtained from the rotation matrix $R(\gamma)$:
\begin{equation}
\begin{pmatrix}
x\\y
\end{pmatrix}
=R(\gamma)
\begin{pmatrix}
x_0\\y_0
\end{pmatrix},
R(\gamma)=
\begin{pmatrix}
\cos\gamma&\sin\gamma\\
-\sin\gamma&\cos\gamma
\end{pmatrix}.
\end{equation}

For the symmetric strain tensor, using $\epsilon=R\epsilon^0R^T$ one can show that
\begin{equation}
\begin{aligned}
\epsilon^0_{xx}-\epsilon^0_{yy}&=\cos2\gamma(\epsilon_{xx}-\epsilon_{yy})-2\sin2\gamma\epsilon_{xy}\\
\epsilon^0_{xy}&=\frac{1}{2}\sin2\gamma(\epsilon_{xx}-\epsilon_{yy})+\cos2\gamma\epsilon_{xy}
\end{aligned}.
\end{equation}

Equivalently,
\begin{equation}
\textbf{A}'=R(-2\gamma)\textbf{A}^0,
\end{equation}
where
\begin{equation}
\textbf{A}'=\frac{\hbar\beta}{2a}\left(\epsilon_{xx}-\epsilon_{yy},-2\epsilon_{xy}\right)^T.
\end{equation}

Plugging the above results together with $\textbf{p}=R(\gamma)\textbf{p}^0$ back into the original Hamiltonian in the $x_0oy_0$ frame, one can verify that the Hamiltonian is given by
\begin{equation}
\mathcal{H}_{\pm}=v_F
\begin{pmatrix}
0&e^{-i\gamma}\left(\pi_x^\mp-i\pi_y^\mp\right)\\
e^{i\gamma}\left(\pi_x^\mp+i\pi_y^\mp\right)&0
\end{pmatrix},
\end{equation}
where $\pi_{x,y}^\mp=p_{x,y}\mp A_{x,y}$, the upper (lower) sign is for valley $K$ ($K'$). The gauge field in the rotated frame reads
\begin{equation}
\textbf{A}=R(3\gamma)\textbf{A}'=R(3\gamma)\frac{\hbar\beta}{2a}\left(\epsilon_{xx}-\epsilon_{yy},-2\epsilon_{xy}\right)^T.
\end{equation}
This gauge field has a $2\pi/3$ periodicity inherited from the 3-fold symmetry of the hexagonal lattice. 

One can also get the above results simply by rotating each vector in $\mathcal{H}^0_\pm=v_F\boldsymbol{\sigma}^0\cdot\left(\textbf{p}^0\mp\textbf{A}^0\right)$:
\begin{equation}
\begin{aligned}
\boldsymbol{\sigma}&=R(\gamma)\boldsymbol{\sigma}^0\\
&=(\cos\gamma\sigma^0_x+\sin\gamma\sigma^0_y,-\sin\gamma\sigma^0_x+\cos\gamma\sigma^0_y)^T
\end{aligned},
\end{equation}
i.e.
\begin{equation}
\sigma_x=
\begin{pmatrix}
0&e^{-i\gamma}\\
e^{i\gamma}&0
\end{pmatrix},
\sigma_y=
\begin{pmatrix}
0&-ie^{-i\gamma}\\
ie^{i\gamma}&0
\end{pmatrix},
\end{equation}
and
\begin{equation}
\textbf{A}=R(\gamma)\textbf{A}^0=R(3\gamma)\textbf{A}',
\end{equation}
where $\textbf{p}^0$ is trivially written as $\textbf{p}$.

If we further perform a unitary transformation
\begin{equation}
U=
\begin{pmatrix}
1&0\\0&e^{-i\gamma}
\end{pmatrix},
\end{equation}
one obtains
\begin{equation}
H_\pm=U\mathcal{H}_{\pm}U^\dagger=v_F\boldsymbol{\sigma}\cdot\left(\textbf{p}\mp\textbf{A}\right),
\end{equation}
which is the same Hamiltonian as the one in the original $x_0oy_0$ frame. As an unitary transformation will not change the transmission probability of the problem, we will employ $H_\pm$ instead of $\mathcal{H}_{\pm}$ in the following.

In the $xoy$ frame, the Gaussian fold is always given by
\begin{equation}
h(\br)=h_0e^{-y^2/b^2},
\end{equation}
which yields
\begin{equation}
\textbf{A}'=\frac{\hbar\beta}{2a}\left(\epsilon_{xx}-\epsilon_{yy},-2\epsilon_{xy}\right)^T=-\frac{\overline{\beta}\eta^2}{v_F}g\left(\frac{y}{b}\right)\left(1,0\right)^T,
\end{equation}
with $\overline{\beta}=\frac{\hbar\beta v_F}{2a}\approx7eV$, $\eta=\frac{h_0}{b}$, and $g(z)=2z^2e^{-2z^2}$. Consequently, the gauge field in the rotated frame reads
\begin{equation}
\textbf{A}=-\frac{\overline{\beta}\eta^2}{v_F}g\left(\frac{y}{b}\right)\left(\cos3\gamma,-\sin3\gamma\right)^T,
\end{equation}
which only depends on the $y$ coordinate, and the fold orientation with respect to the zigzag direction $\gamma$.

In order to solve the Schr\"odinger equation:
\begin{equation}
\begin{aligned}
H_\tau\Psi^\tau(x,y)&=v_F\boldsymbol{\sigma}\cdot\left(\textbf{p}-\tau\textbf{A}\right)\Psi^\tau(x,y)\\
&=E\Psi^\tau(x,y)
\end{aligned},
\end{equation}
we propose
\begin{equation}
\Psi^\tau(x,y)=\psi^\tau(y)e^{ik_xx}=
\begin{pmatrix}
\psi_{I}^\tau(y)\\\psi_{II}^\tau(y)
\end{pmatrix}
e^{ik_xx}
\end{equation}
Using this ansatz, we arrive at
\begin{equation}
\begin{aligned}
&\begin{pmatrix}
0&\hbar k_x-\tau A_x-i\pi_y^{-\tau} \\
\hbar k_x-\tau A_x+i\pi_y^{-\tau}&0
\end{pmatrix}
\begin{pmatrix}
\psi_{I}^\tau(y)\\\psi_{II}^\tau(y)
\end{pmatrix}\\
&=\frac{E}{v_F}
\begin{pmatrix}
\psi_{I}^\tau(y)\\\psi_{II}^\tau(y)
\end{pmatrix}
\end{aligned}.
\end{equation}
This yields two coupled differential equations, which in general cannot be solved analytically. We will separate $A_x$ and $A_y$ into narrow slices in the $y$ coordinate, where in each slice they can be taken as constants. In the $i$-th slice, keep in mind that $A_{x,i}$ and $A_{y,i}$ are independent of $y$, one can verify that the two coupled equations can be casted into
\begin{equation}
\left[\left(k_x-\tau \frac{A_{x,i}}{\hbar}\right)^2+\left(\frac{p_y}{\hbar}-\tau \frac{A_{y,i}}{\hbar}\right)^2\right]\psi_i^\tau(y)=k^2\psi_i^\tau(y),
\end{equation}
where $k^2=\frac{E^2}{\hbar^2v_F^2}$. Rearranging the equation, 
\begin{equation}
\left(\frac{d^2}{dy^2}-\tau2i\frac{A_{y,i}}{\hbar}\frac{d}{dy}-\frac{A_{y,i}^2}{\hbar^2}+q_{\tau,i}^2\right)\psi_i^\tau(y)=0,
\end{equation}
where
\begin{equation}
q_{\tau,i}=\sqrt{k^2-\left(k_x-\tau \frac{A_{x,i}}{\hbar}\right)^2}.
\end{equation}
The general solution to the equation is $\psi_i^\tau(y)=(1,c_i^\tau)^Te^{\lambda_{\tau,i} y}$, where $\lambda_{\tau,i}$ satisfies
\begin{equation}
\lambda_{\tau,i}^2-\tau2i\frac{A_{y,i}}{\hbar}\lambda_{\tau,i}-\frac{A_{y,i}^2}{\hbar^2}+q_{\tau,i}^2=0,
\end{equation}
i.e. 
\begin{equation}
\lambda_{\tau,i}^\pm=i\left(\tau\frac{A_{y,i}}{\hbar}\pm q_{\tau,i}\right).
\end{equation}
By replacing this solution back in the Schr\"odinger equation, one obtains
\begin{equation}
c_i^\tau=\frac{1}{E}\left(\hbar v_F k_x-\tau v_F A_{x,i}\pm i \hbar v_F q_{\tau,i}\right).
\end{equation}
In summary, in the $i$-th slice, one has 
\begin{equation}
\begin{aligned}
\psi^\tau_i(y)
&=t_i^\tau
\begin{pmatrix}
1\\\frac{\hbar v_F k_x-\tau v_F A_{x,i}+i \hbar v_F q_{\tau,i}}{E}
\end{pmatrix}
e^{i\tau \frac{A_{y,i}}{\hbar}y}e^{iq_{\tau,i}y}\\
&+r_i^\tau
\begin{pmatrix}
1\\\frac{\hbar v_F k_x-\tau v_F A_{x,i}-i \hbar v_F q_{\tau,i}}{E}
\end{pmatrix}
e^{i\tau \frac{A_{y,i}}{\hbar}y}e^{-iq_{\tau,i}y}
\end{aligned}
\end{equation}

These expressions reveal that $A_y(y)$ does not contribute to the pseudo-magnetic, and it can be removed from the Hamiltonian by a gauge transformation $\Psi^{\tau}=e^{i\tau F(y)/\hbar}\tilde{\Psi}^{\tau}$, where $F(y)=\int^{y}A_y(y')dy'$, or equivalently $\frac{dF(y)}{dy}=A_y(y)$. In other words, $v_F\boldsymbol{\sigma}\cdot\left(\textbf{p}-\tau\textbf{A}\right)\tilde{\Psi}^\tau(x,y)
=E\tilde{\Psi}^\tau(x,y)$ is satisfied, with $\textbf{A}=-\frac{\overline{\beta}\eta^2}{v_F}g\left(\frac{y}{b}\right)\left(\cos3\gamma,0\right)^T$. If $\textbf{A}$ is assumed to be constant, we recover the result given above in each slice of the gauge field. 

The above discussions focus on the pseudo-vector potential, as to the scalar potential, one can easily check that $\epsilon_{xx}+\epsilon_{yy}$ is not affected by the rotation, so the scalar potential remains the same. 

\bibliography{RefsValley}

\end{document}